\documentstyle[12pt,epsf]{article}
\def\pr{Phys.\ Rev.\ }
\def\prl{Phys.\ Rev.\ Lett.\ }
\def\pl{Phys.\ Lett.\ }
\def\np{Nucl.\ Phys.\ }
\def\zp{Z.\ Phys.\ }

\def\hep{hep-ph/96}
\def\rf#1#2#3{{\bf #1}, #2, (19#3)}
\def\etal{{\it et. al.}}
\def\VEV#1{\left\langle #1\right\rangle}
\renewcommand{\thefootnote}{\fnsymbol{footnote}}

\newcommand{\fcaption}[1]{
        \refstepcounter{figure}
        \setbox\@tempboxa = \hbox{\footnotesize Fig.~\thefigure. #1}
        \ifdim \wd\@tempboxa > 6in
           {\begin{center}
        \parbox{6in}{\footnotesize\baselineskip=12pt Fig.~\thefigure. #1}
            \end{center}}
        \else
             {\begin{center}
             {\footnotesize Fig.~\thefigure. #1}
              \end{center}}
        \fi}
\begin{document}
\begin{titlepage} \rightline{\vbox{\halign{#\hfil \cr
   \normalsize ANL-HEP-PR-96-103 \cr
   \normalsize hep-ph/97xxyyy \cr
   \normalsize\today \cr}}}
\vspace{.75in} 

\begin{center} 
\Large 
{\bf  Probing Nucleon Spin Structure}
\footnote{Submitted for publication in ``Progress in Particle and Nuclear
Physics", Volume 39. Work supported by the U.S. Department of Energy,
Division of High Energy Physics, Contract W-31-109-ENG-38.
e-mail: GPR@hep.anl.gov} 
\medskip\medskip\medskip

\normalsize Gordon P. Ramsey
\\
\smallskip
{\it Physics Department, Loyola University of Chicago, Chicago, IL 60626}
\\
\smallskip
and
\\
\smallskip
{\it High Energy Physics Division, Argonne National Laboratory, IL 60439}
\end{center}

\begin{abstract}
One of the important questions in high energy physics is the relation of quark
and gluon spin to that of the nucleons which they comprise. Polarization
experiments provide a mechanism to probe the spin properties of elementary
particles and provide crucial tests of Quantum Chromodynamics (QCD).
The theoretical and experimental status of this fundamental question will be
reviewed in this paper.
\end{abstract}

\renewcommand{\thefootnote}{\arabic{footnote}} \end{titlepage}

\section{INTRODUCTION}

\subsection{Motivation for Spin Studies}

In recent years there has been a significant interest in the study of spin
phenomena. It is now widely accepted that the physics of spin phenomena in
particle interactions provides vital information on the most profound
properties of particles: their wave functions, the short and long distance
dynamics of the quark and gluon interactions, and the mechanisms of chiral
symmetry breaking and confinement. These systematic spin studies have several
well defined goals; some of these include:
\begin{itemize}
\vspace*{-2ex}
\item
The study of the spin structure of the nucleon, i.e., how the proton's spin
state can be obtained from a superposition of Fock states with different
numbers of constituents with spin.
\vspace*{-2ex}
\item
How the dynamics of constituent interactions depend on spin.
\vspace*{-2ex}
\item
Study the overall nucleon structure and long range dynamics.
\vspace*{-2ex}
\item
Understand chiral symmetry breaking and helicity non-conservation on the
hadronic level.
\end{itemize}
\vspace*{-2ex}
The first two goals are related to the constituent short-interaction dynamics,
while the last two concentrate on hadronic long-range (non-perturbative)
dynamics. Experiments designed to investigate all of these important properties
of nucleons can be performed with polarized proton beams at many present
accelerators. These experiments can also be used to test some fundamental
assumptions of QCD regarding spin. \\

The naive quark model has been successful in predicting most of the gross
properties of hadrons, such as charge, parity, isospin and symmetry properties
in relation to each other. Some of the dynamics of particle interactions can
qualitatively be understood in terms of this model, as well. However, it has
been noted, within the past decade, that it falls short in explaining the
spin properties of hadrons in terms of their constituents. Perturbative QCD
(PQCD) has also been successful in predicting asymptotic properties of hadronic
dynamics in the limit of short range interactions. Once again, some of the
PQCD predictions regarding spin have disagreed with data, even in those
kinematic regions where PQCD is thought to be valid. The data include
measurements of analyzing power, which yielded non-zero values, in contrast to 
PQCD predictions and oscillations in exclusive $p-p$ scattering cross sections.
These are related to the assumption of helicity conservation on the hadronic
level. Hyperon production experiments from inclusive $p-p$ scattering have
shown that polarized hyperons can be produced from unpolarized protons. This
is totally unexpected from PQCD arguments and the assumption of the pointlike
constituent dynamics. These phenomena are likely related to longer range
dynamics and the coordinated rotation of constituents in terms of angular
momentum. \\

Thus, in probing hadronic structure with regard to spin, a number of
interesting events have arisen, which require modification of the models
created to explain hadronic matter. Outside of new particle discoveries,
spin phenomena have provided the high energy community with a great number of
surprises, which have forced us to refine our picture of elementary matter. \\

In this review, I will discuss a particular aspect of this exciting field,
namely the quest for understanding nucleon spin with regard to the quark and
gluon constituents and their overall collective motion. Naturally, there is
some overlap in the understanding of the constituent spin and the overall
nucleon spin, since the constituents must be confined to the nucleon ``bag".
Also, since the collective motion of the constituents plays and important role
in structure and the dynamics of the interactions, some of the analyses
will include the non-perturbative effects which are characteristic of the long
range dynamics. The main thrust of this paper, however, is to describe the
theoretical and experimental aspects of understanding the constituent
structure of nucleon spin in fair detail. \\

\subsection{The Spin Crisis}

Consituent spin structure studies involve inclusive processes such as the deep
inelastic scattering (DIS) of leptons on nucleons. There are two crucial reasons
why deep-inelastic scattering is the primary tool for probing hadronic
structure: (1) the pointlike nature of the leptons allows the probing of
hadrons so that constituent interactions can be ignored, and (2) the
theoretically calculable quantities which characterize the constituent
structure can be directly related to the measurable cross sections. In other
words, the cross sections are factorizable into a calculable part and a
measured part. Historically, the deep-inelastic scattering of electrons first
proved the existence of point-like constituents in the nucleon. The first spin
structure measurements of the proton were performed in the 1970's when the
SLAC-Yale collaboration measured the spin structure function $g_1(x)$.
\cite{Alguard} In the late 1980's, the European Muon Collaboration at CERN
\cite{EMC} extended the kinematic range of the DIS measurements using a
polarized muon beam. The extrapolation of the polarized structure function
$g_1^p$ to lower $x$ (parton fractional momentum) led to implications that, 
although the Bjorken sum rule of QCD \cite{Bjorken66} was satisfied, the
Ellis-Jaffe sum rule, \cite{EJSR} based on a simple quark model with unpolarized
strange quarks, was violated. This created a controversy, labeled the
``spin crisis" which heightened the interest in spin phenomena. Since then,
a flurry of theoretical and experimental work has been performed to address
this ``crisis" and further investigate the spin properties of the lighter
hadrons. \\

Initial analyses of the EMC data were motivated by comparison to the naive
quark model and the Ellis-Jaffe sum rule. This sum rule was based on the
assumption of an unpolarized strange sea, motivated by unitarity arguments,
and will be discussed in detail later. The ``spin-crisis" arose since the
data disagreed considerably with the naive quark model, which predicted that
all of the nucleon spin should be carried by the valence quarks, and the
Ellis-Jaffe sum rule. The EMC data led to the implications that:

\begin{itemize}
\vspace*{-2ex}
\item (1) the total quark content of the proton spin is small and possibly
consistent with zero, \\
\vspace*{-2ex}
\item (2) the predictions of the E-J sum rule are severely violated.
\end{itemize}

There were a number of proposed ``fixes" for these problems. \cite{EMCTh} Among
them were: \\

\begin{itemize}
\vspace*{-2ex}
\item (1) introduction of the gluon anomaly, which can modify the spin content
of each quark flavor, so that the measured and calculated values differ by
a calculable factor, \\
\vspace*{-2ex}
\item (2) proposing that the spin carried by the polarized gluons was extremely
large ($\VEV{\Delta G}\approx 6$), \\
\vspace*{-2ex}
\item (3) assuming that the corresponding orbital motion of the constituents was
very large, \\
\vspace*{-2ex}
\item (4) assuming that the polarized sea was large and negative, canceling
most of the valence quark contribution to the spin, \\
\vspace*{-2ex}
\item (5) abandonment of the quark model in favor of alternatives, such as the
Skyrme model.
\end{itemize}

All of these were motivated by the prevailing notion that the naive quark
model with an unpolarized strange sea was an accurate model of the nucleons'
spins. Analyses of the more recent data (from 1992 on) are more inclined to
accept the notion that the naive quark model must be modified to account for
spin contributions of the constituents and that the sea is polarized negatively
with respect to the valence quarks. This virtually eliminates the necessity
of making the polarized gluons or the angular momentum extremely large. There
may be other considerations which affect these quantities in the opposite
way, such as negatively polarized gluons and the possibility of rapidly growing
structure functions at small $x$. Thus, in response to the recent DIS data,
previling theories have reduced the possible range of values of the
constituents' spin contributions and have made us reconsider the important
elements which modify these contributions. \\

Recently, much attention of the HEP community has focused on the physics of
polarized hadron interactions. In the last couple of years, experimental groups
at SLAC, CERN and DESY have improved statistics and lowered the systematic
errors in these DIS experiments. \cite{SMC, E142, HERMES}
This data has allowed us to draw several conclusions about the nucleon spin
structure in the kinematic region of lepton momentum transfer, $Q^2\simeq 2\to
10$ GeV$^2$ and Bjorken $x\simeq 0.002\to 0.7$. The analysis of these data in
the framework of perturbative QCD provides information on longitudinally
polarized parton distributions $\Delta q_i(x,Q^2)$, interpreted as the
differences of probabilities $q^{+/-}(x,Q^2)$ for finding partons of the type
$i$ with spin parallel/antiparallel to the spin of the parent nucleon. The
results have required modification of the earlier quark models of nucleon spin
and have created other controversies regarding the amount of spin carried by
the strange sea quarks and the gluons. \cite{GR95} Thus, although there has
been significant progress in understanding nucleon spin structure, present 
data have left unanswered questions regarding: (1) flavor dependent spin
structure, (2) relations between the polarized and unpolarized quark and gluon
structure functions and (3) choice of factorization of the quark distributions
and the role of the gluon anomaly in their determination.

Recently, the E704 group at Fermilab has analyzed the data from their $pp$
and $p\bar p$ experiments for jet production cross sections. \cite{E704}
The results have implied that the polarized gluon distribution is limited
in size, but has not put a strict value on this limit. In all of the inclusive
experiments done this far, we have gained some knowledge of the constituent
spin content, but it is clear that much more work has to be done in future
experiments to obtain a complete understanding of this important problem in
physics. Since the initial studies of constituent spin structure have relied
heavily upon deep-inelastic scattering experiments, it is useful to outline the
connection between theoretical calculations and the measured cross sections. \\
\\

\subsection{Deep-Inelastic Scattering} 

\subsubsection{Formalism}

Since the pointlike lepton is incident on the hadron target at very high
energy, it effectively strikes one of the quarks within the hadron. From the
angular distribution of the scattered lepton, the momentum information can be
inferred. The polarized deep-inelastic scattering process (DIS) can be
represented by the diagram in Fig. 1. 

\begin{figure}
{\hskip 5.0cm}\hbox{\epsfxsize9.0cm\epsffile{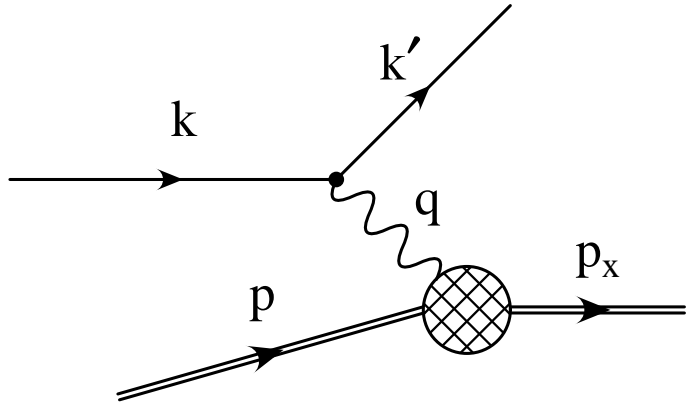}}
\caption{Diagram of Deep-Inelastic Scattering, showing relavent momenta.}
\end{figure}

In this figure, the initial proton (or hadron target) and lepton $4-$momenta
are given by $p^\mu$ and $k^\mu$, respectively. The virtual exchanged photon
has momentum  $q$ and the outgoing struck quark and lepton momenta are
$p_x^\mu$ and $k^{\mu \prime}$, respectively. The cross sections are usually
written in terms of the invariants: $Q^2=-q^2$, $\nu=E-E^{\prime}$ and
$x=Q^2/(2M\nu)$.
Here, $Q^2$ is the momentum (squared) transfered to the nucleon by the lepton,
$\nu$ is the energy lost by the lepton in the collision and thus transferred by
the photon to the quark, and $x$ is a dimensionless variable, called the
Bjorken scaling variable, which is a measure of the fraction of momentum
carried off by the struck quark. \cite{Bjorken69} Cross section information and
the corresponding theoretically generated structure functions are represented
in terms of either the variables $\nu$ and $Q^2$ or $x$ and $Q^2$.

Let us briefly review the formalism of deep inelastic polarized electron
scattering on a polarized proton. \cite{Roberts}
Since the thrust of this review is to study nucleon spin structure via the
constituents, we will cover only the distribution part of the interaction,
which involves the structure functions. The final-state fragmentation functions
are not considered here, but the reader can refer to recent treatments of
spin measurements of the fragmentation functions done by Anselmino, \etal,
\cite{Anselmino96a} and Mulders. \cite{Mulders}

In the distribution function case, the hadronic tensor carries all
of the relavent information about the target nucleon. This tensor, labeled
$W_{\mu \nu }$, will depend on the momenta $p$ and $q$ and on the covariant
pseudovector of the proton spin, $s$. Taking into account translation
invariance, the hadronic tensor can be rewritten as a Fourier transformation
of the matrix element of the commutator of two currents representing the
dynamics of the interaction:
\begin{equation}
W_{\mu \nu}(p,q,s)=\frac{1}{4M}\int \frac{d^4\xi}{2\pi}\>e^{iq\xi}\langle p,s|
[J_\mu(\xi),J_\nu(0)]|p,s\rangle=W_{\mu \nu}^{[S]}+iW_{\mu \nu}^{[A]},
\label{1.1}
\end{equation}
where $|p,s\rangle$ denotes the proton state with momentum $p$ and spin $s$.
The hadronic tensor contains all of the appropriate information necessary to
extract the spin-averaged (unpolarized) and spin-weighted (polarized) structure
functions in inclusive scattering. For the purpose of separating the
unpolarized and polarized information, we can separate the symmetric [S] and
anti-symmetric [A] parts of the hadronic tensor. The functions $W_{\mu \nu}
^{[S]}$ and $W_{\mu \nu}^{[A]}$ are determined by the discontinuity of the
symmetric and antisymmetric parts of the forward Compton scattering  amplitude:
$W_{\mu \nu }^{[S,A]}=\frac{1}{\pi }\mbox{Im}\>T_{\mu \nu }^{[S,A]}$.
Imposing parity (P), charge conjugation (C) and current conservation ($q^\mu
W_{\mu\nu}=0$) we can extract the forms:
\begin{equation}
W_{\mu \nu }^{[S]}=(-g_{\mu \nu}-{{q_\mu q_\nu}\over {Q^2}})W_1(x,Q^2)+
\frac{1}{M^2}(p_{\mu}+{{p\cdot q}\over {Q^2}}q_{\mu})(p_{\nu}+{{p\cdot q}\over
{Q^2}}q_{\nu})W_2(x,Q^2) \label{1.2}
\end{equation}
and
\begin{equation}
W_{\mu \nu }^{[A]}=\frac{1}{M^4}\varepsilon _{\mu \nu \lambda \sigma }q^\lambda
[M^2s^\sigma G_1(\nu,Q^2)+(Mp\cdot q s^\sigma -s\cdot q p^\sigma )G_2(\nu,Q^2)].
\label{1.3}
\end{equation}
This last relation provides the opportunity to apply the helicity amplitude
formalism which is especially useful in the case of targets with spins greater
than $1\over 2$. For a spin-$1\over 2$ target there are only four independent
helicity amplitudes in forward Compton scattering and hence four independent
structure functions. The spin-dependent structure functions $G_1$ and $G_2$
determine the asymmetries which depend on the initial state lepton and nucleon
polarizations. 

In the Bjorken limit ($\nu$ and $Q^2$ $\to \infty$, $x$ constant) scaling is
valid and the functions $W_1$, $W_2$, $G_1$ and $G_2$ should depend only on
$x$, up to the logarithmic corrections:
\begin{eqnarray}
MW_1(\nu ,Q^2)\rightarrow F_1(x), \nonumber \\
\nu W_2(\nu ,Q^2)\rightarrow F_2(x),\nonumber \\
M^2\nu G_1(\nu ,Q^2)\rightarrow g_1(x), \nonumber \\
M\nu ^2G_2(\nu ,Q^2)\rightarrow g_2(x).\label{1.4}
\end{eqnarray}
Studies of deep inelastic scattering processes in the Bjorken scaling limit
provide us with the knowledge on the structure of the target nucleon.
In the Bjorken scaling limit, the symmetric and anti-symmetric parts of the
hadronic tensor are:
\begin{eqnarray}
&   &
W_{\mu \nu}^{[S]}=\frac{1}{M}(-g_{\mu \nu}+{{q_\mu q_\nu}\over {q^2}})
F_1(x,Q^2)+\frac{1}{M^2\nu}(p_{\mu}+{{p\cdot q}\over {Q^2}}q_{\mu})(p_{\nu}
+{{p\cdot q}\over {Q^2}}q_{\nu})F_2(x,Q^2) \nonumber \\
&   &
W_{\mu \nu}^{[A]}={{\varepsilon_{\mu \nu \lambda \sigma}q^\lambda s^\sigma}
\over {M^4\nu}} g_1(x,Q^2)+{{\varepsilon_{\mu \nu \lambda \sigma }(M\nu s^\sigma
-q\cdot s p^{\sigma}})\over {M\nu^2}}g_2(x,Q^2). \label{1.5}
\end{eqnarray}
The functions $F_1$ and $F_2$ are the structure functions extracted from
unpolarized DIS, while $g_1$ and $g_2$ are extracted from polarized DIS.
These reveal the pointlike interaction between the hard photon and the
constituent partons within the hadron.

In the operator product expansion of QCD, the local operators are quark and
gluon operators, characterized by definite values of twist. The hadronic
electromagnetic current in the quark model has the form:
\begin{equation}
J_\mu (\xi )=\sum_q e_q^2\bar{q}(\xi )\gamma _\mu q(\xi ). \label{1.6}
\end{equation}
Here  $q(\xi)$ denotes the quark field. The matrix element between the
free quark states is
\begin{equation}
\langle q|JJ|q \rangle \sim c_q\langle q|O_q|q \rangle+ c_g\langle q|O_g|q
\rangle, \label{1.8}
\end{equation}
where indices $q$ and $g$ refer to the quark and gluon operators.
Since the electromagnetic current is a quark operator the left hand side and
first term in the right hand side are of order $(\alpha _s)^0$, whereas the
matrix element $\langle q|O_g|q \rangle$ is of order $(\alpha _s)^1$, because
there are at least two gluons entering the operator $O_g$. Thus, in general
one can neglect the gluon operator. The most important operators in OPE are the
operators with the lowest possible twist. The twist-two operators
correspond to the vector and axial vector currents for the process of the
virtual Compton scattering which determine the hadronic tensor $W_{\mu \nu }$.
These operators provide the finite contribution to the structure functions
in the deep inelastic limit while the contribution of twist-three operators
are suppressed by $M/Q$. Transverse spin structure functions, for example,
are of the twist-three type.

The structure function $g_1(x,Q^2)$ is related to the longitudinal polarization
of the proton spin with respect to its momentum, i.e. its {\it helicity}. In the
parton model, $g_1$ effectively measures the quark helicity density.
The second spin-dependent proton structure function $g_2(x,Q^2)$ is related
to transverse polarization of the nucleon spin. It has been measured at SLAC
(E143) in a limited $x$ region. The analysis based on the OPE does not depend on
the type of nucleon polarization (longitudinal or transverse). The analysis
of the second structure function $g_{2}(x,Q^2)$ may be performed similar to
the function $g_{1}(x,Q^2)$. The significant difference between the two cases
is that $g_1(x,Q^2)$ receives a contribution only from twist-two operators
whereas $g_2(x,Q^2)$ gets contributions from both twist-two and twist-three
operators simultaneously. A few theoretical results have been obtained
for the function $g_2(x)$, but simple partonic interpretation is only possible
for the twist-two operator contribution. The twist-three contributions are not
well understood. The following relation exists between the functions $g_1$ and
the twist-two contribution to $g_2$: \cite{WW}
\begin{equation}
g_2(x,Q^2)=\int_x^1\frac{dy}{y}g_1(y,Q^2)-g_1(x,Q^2).\label{1.8b}
\end{equation}
The above relation is used to calculate $g_2(x,Q^2)$ from $g_1(x',Q^2)$
at $x'\geq x$ in the framework of the parton model with free on-shell partons.
However there are no reasons to neglect the contributions of twist-three
operators at low enough $Q^2$ and therefore $g_2$ can be represented as follows:
\begin{equation}
g_2(x,Q^2)=g_2(x,Q^2)^{[2]}+g_2(x,Q^2)^{[3]},\label{1.9}
\end{equation}
where the first term on the right hand side of Eqn. (9) is provided by Eqn. (8).
The twist-three operator contributions $g_2(x,Q^2)^{[3]}$
depend on the effects of quark-gluon interactions and quark masses. \\
                                     
Experimental measurements of $g_2(x,Q^2)$ provide a direct way to study the
magnitude of twist-3 contributions and accurate data with polarized proton
beams would be crucial in the resolving of the question of the magnitude of
these contributions. Data could also provide a test for the
Burkhardt-Cottingham sum rule: \cite{BC}
\begin{equation}
\int_0^1 g_2(x)=0.\label{1.10}
\end{equation}
The success or failure of Eq. 1.10 depends on its long-range behavior
which, in the simple models, is such that this sum rule is satisfied in
perturbative QCD. Thus, valuable information regarding both perturbative and
non-perturbative processes can be obtained by testing this sum rule.
Recent measurements of $g_2$ by the SMC \cite{SMC} and E143 \cite{E143}
experiments have verified the condition that $A_2<\surd R$ ($R \equiv \sigma_L
/\sigma_T$) so that the leading twist terms dominate and that the
Burkhardt-Cottingham sum rule is valid to within experimental errors.
There are a number of recent treatments of the role of $g_2$ in DIS with regard
to transverse spin content and chiral symmetry breaking. \cite{Tspin}
This topic will not be discussed here. \\

The above discussion has shown how the hadronic tensor can be written in terms
of the spin-averaged (unpolarized) structure functions $F_1$ and $F_2$, and the
spin-weighted (polarized) structure functions $g_1$, $g_2$. It is these
quantities which can be extracted from the cross sections in various DIS
processes. The phenomenology of DIS starts with combining the incident lepton
tensor with the hadronic tensor and relating this product to the measured cross
sections. \\

\subsubsection{Phenomenology}

The polarized lepton tensor $L_{\mu\nu}$, being pointlike, can be written
explicitly as:
\begin{equation}
L_{\mu\nu}=2\Bigl[k_\mu k'_\nu+k'_\mu k_\nu-g_{\mu\nu}k\cdot k'-m_{\mu}^2)+
i\varepsilon_{\mu\nu\rho\sigma}q^{\rho}s^{\sigma}\Bigr], \label{1.11}
\end{equation}
where the antisymmetric last term is absent in the unpolarized case.
The product of $L_{\mu\nu}$ and $W_{\mu\nu}$ is directly related to the cross
sections measured in the process of DIS.

Since cross sections are the most fundamental measured property of elementary
particles, all of the theoretical quantities characterizing spin must be
extracted from some combination of the polarization cross sections.
\cite{Roberts} The following differential cross section relates to the lepton
and hadron tensor product as:
\begin{equation}
{{d^2\sigma} \over{d\nu dQ^2}}={{4\pi \alpha M^2} \over{(s-M^2)^2\>Q^4}}
\cdot L_{\mu \nu}W^{\mu \nu}. \label{1.12}
\end{equation}
This connects the structure functions in the Bjorken limit directly to the
measured cross sections, both in the unpolarized and polarized cases. 
These differential cross sections can be written as  in terms of any of the
kinematic variables. For example, if the energies and angular distributions
of the leptons are measured, the unpolarized and polarized structure functions
are related to the spin averaged and spin weighted differential cross sections
by:
\begin{equation}
{{d^2\sigma(\rightarrow \rightarrow)} \over{d\Omega dE'}}+
{{d^2\sigma(\leftarrow \rightarrow)} \over{d\Omega dE'}}={{8\alpha^2 E'^2}
\over {MQ^4}}\Bigl[2\sin^2(\theta/2)F_1(x,Q^2)+(M^2/\nu)\cos^2(\theta/2)F_2
(x,Q^2)\Bigr], \label{1.13}
\end{equation}
and
\begin{equation}
{{d^2\sigma(\rightarrow \rightarrow)} \over{d\Omega dE'}}-
{{d^2\sigma(\leftarrow \rightarrow)} \over{d\Omega dE'}}={{4\alpha^2 E'}
\over {Q^2E\nu}}\Bigl[(E+E'\cos \bigl(\theta)\bigr)g_1(x,Q^2)-(2xM)g_2(x,Q^2)
\Bigr], \label{1.14}
\end{equation}
where $d\Omega$ is the angular distribution of the leptons.

In nature, asymmetric behavior often occurs between the interactions of
particles whose spins are aligned and interactions of particles whose spins are
anti-aligned. Thus, it is convenient to define an asymmetry as:
$$
A=\Biggl[{{\sigma(++)-\sigma(+-)}\over {\sigma(++)+\sigma(+-)}}\Biggr],
\label{1.15}
$$
where the ++ refers to the aligned spins and $+-$ to the anti-aligned spins,
without regard to the direction of polarization relative to the momentum of
the interaction. There are distinct advantages to defining the asymmetries in
this way. Theoretically, $A$ is a ratio of cross sections, so all of the
normalizations used to calculate cross sections cancel, leaving only the key
parameters of the interaction. Experimentally, the difference in the numerator
tends to cancel some of the systematic errors, making $A$ a fundamentally
more sensitive measure of the physical variables.

There are asymmetries for both the longitudinal and transverse spin relative
polarizations. The lepton-hadron asymmetries are given in terms of the
cross-section as:
\begin{eqnarray}
&   &
A_{\parallel}(x,Q^2)=\Biggl[{{\sigma(\leftarrow \rightarrow)-\sigma(\leftarrow
\leftarrow)}\over {\sigma(\leftarrow \rightarrow)+\sigma(\leftarrow \leftarrow)
}}\Biggr] \equiv \biggl[{{\Delta \sigma_{parallel}}\over {\sigma}}\biggr]
\nonumber \\
&   &
A_{\perp}(x,Q^2)=\Biggl[{{\sigma(\uparrow \downarrow)-\sigma(\uparrow \uparrow)}
\over {\sigma(\uparrow \downarrow)+\sigma(\uparrow \uparrow)}}\Biggr]
\equiv \biggl[{{\Delta \sigma_{perp}}\over {\sigma}}\biggr], \label{1.16}
\end{eqnarray}
where the arrows refer to the relative longitudinal spin directions of the
beam and target, respectively.

The structure functions can be related to the absorptive cross sections,
$\sigma_{1\over 2}$ and $\sigma_{3\over 2}$ for virtual photons with helicity
projections $1\over 2$ and $3\over 2$, respectively. The asymmetry components
$A_1$ and $A_2$ can be written in terms of the corresponding cross sections
along with the interference between the longitudinal and transverse
polarizations, $\sigma_I$. The parallel asymmetry in terms of these components
can be written as
\begin{eqnarray}
&  &
A_{\parallel}=D(A_1+\eta A_2), \nonumber \\
&  &
   =D\Bigl[{{\sigma_{1\over 2}-\sigma_{3\over 2}}\over {\sigma_{1\over 2}+
\sigma_{3\over 2}}}\Bigr]+ 2D\eta{{\sigma_I}\over {\sigma_{1\over 2}+
\sigma_{3\over 2}}}.  \label{1.17}
\end{eqnarray}
The kinematic factors are: $D={{1-(1-y)\epsilon}\over {1+\epsilon R}}$ and
$\eta={{2M\epsilon(Q^2)^{1\over 2}}\over{s[1-(1-y)\epsilon]}}$, where $y$ is
the fraction of energy lost be the lepton in the lab frame and
$R=\sigma_L/\sigma_T$, the ratio of transverse to longitudinal photon-nucleon
($\gamma-N$) cross sections. In most DIS experiments, the kinematics are such
that $D$ is known for each experiment and $\eta \sim \epsilon Q/s$ is very
small. The asymmetries can be linked to the structure functions $g_1$ and
$g_2$ by the following sequence of relations. First, the longitudinal asymmetry
is factored into a combination of the longitudinal $\gamma-N$ asymmetry,
$A_1(x,Q^2)$ and the transverse $\gamma-N$ asymmetry, $A_2(x,Q^2)$ in the
following way:
\begin{equation}
A_{\parallel}=D(A_1+\eta A_2)=D\Bigl[{{1+\gamma^2}\over {F_1}}(g_1-\gamma^2\>
g_2)+\eta {{\gamma(1+\gamma^2)}\over {F_1}}(g_1+g_2)\Bigr], \label{1.18}
\end{equation}
where $\gamma^2=(2Mx)^2/Q^2$. From recent measurements, it has been shown that
$A_2$ is small [E143, 1995 and $\gamma$ is also negligible for the SMC
experiments. In addition, it is also assumed that the transverse asymmetry
$A_2$ is small, since it is bounded by $\surd R$. As a result, the structure
function $g_2^p$ is neglected in both factors of $A_{\parallel}$. Experimental
measurements appear to imply that $A_1$ is relatively independent of $Q^2$, an
assumption which has been substantiated by phenomenological analysis of the
data. \cite{Kotikov} However, Gl\"{u}ck, \etal, \cite{Gluck95} point out that
this may not strictly be a valid approximation at low-$x$ and moderate $Q^2$.
Assuming that this is a reasonable approximation, then, in the Bjorken limit,
$F_1=2xF_2$, and thus, $A_1\approx A_{\parallel}/D$ with
\begin{equation}
g_1(x,Q^2)\approx F_1(x,Q^2)\cdot A_1(x)=\Biggl[{{F_2(x,Q^2)\>A_1(x)}\over
{2x(1+R)}}\Biggr]. \label{1.19}
\end{equation}
Information about the polarized quark distributions can be extracted directly
from this asymmetry by
\begin{equation}
A_1(x)={{\sum_{\scriptstyle i} e_i^2\Delta q_i(x)}\over {\sum_{\scriptstyle i}
e_i^2 q_i(x)}}, \label{1.20}
\end{equation}
where $e_i$ is the charge of each quark flavor and for each flavor, $i$:
\begin{equation}
\Delta q_i(x,Q^2)\equiv q_i^+(x,Q^2)-q_i^-(x,Q^2). \label{1.21}
\end{equation}                                        

Using the OPE it can be shown that the first moment of the proton
structure function $g_1^p(x,Q^2)$ is determined by the following matrix
elements of the axial-vector current:
\begin{eqnarray}
\int_0^\infty dx\>g_1^p(x,Q^2)=\frac{1}{2}\left[ \frac{4}{9}\Delta u(Q^2)+
\frac{1}{9}\Delta d(Q^2)+\frac{1}{9}\Delta s(Q^2)\right]\times\nonumber \\
 \left(1-\frac{\alpha _s(Q^2)}{\pi }+O(\alpha _s^2)\right)+
O\left(\frac{\Lambda ^2}{Q^2}\right), \label{1.22}
\end{eqnarray}
where
\begin{equation}
\Delta q(\mu^2)s_\nu=\langle p,s|(\bar{q}\gamma_\nu\gamma_5q){\mid}_{\mu^2}
|p,s\rangle.\label{1.23}
\end{equation}
The term $\mu^2$ is the relevant mass scale or the renormalization point for
the axial-vector current operator. The functions $\Delta q(Q^2)$ are related
to $\Delta q(\mu^2)$ by the QCD evolution equations that will be discussed
later. Note that another leading twist operator, the vector current
$\bar{q}\gamma _\mu q$, provides the finite contribution to the symmetrical
part of the hadronic tensor $W_{\mu \nu}^{[S]}$.

Combining the form of Eqn. 21 for the proton and neutron, we arrive at
the polarized version of the Bjorken sum rule, which relates the first moment 
of the difference between the proton and neutron structure functions,
$g_1^p-g_1^n$ to nucleon beta decay:
\begin{equation}
\int_0^1 [g_1^p(x)-g_1^n(x)]\>dx=\frac{1}{6}\int_0^1 [\Delta u_{total}(x)
-\Delta d_{total}(x)]\>dx=\frac{1}{6}g_A(1-\frac{\alpha_s(Q^2)}{\pi} +h.o.c.),
\end{equation}
where $g_A$ is measured in nucleon beta decay and $h.o.c.$ refer to calculated
higher order QCD corrections. This and other sum rules will be discussed later.
\\

The higher order corrections to DIS structure functions have been analyzed
by various groups. In particular, the higher twist corrections to the proton
and neutron functions by Stein, \etal, \cite{HT} and were found to be $Q^2$
dependent, but very small, even at the lower $Q^2$ values of the data. The
higher twist corrections to the BSR were consistent with zero. Meyer-Hermann,
\etal, calculated the twist-4 contributions to $g_1$ using the IR-renormalon
method \cite{MH} and obtained a result of $\pm 0.017$ GeV$^2/Q^2$. (See also
Boer and Tangerman \cite{BT})
The twist-2 and twist-3 contributions to the BSR are also suppressed by a
factor of $1/Q^2$. The higher order QCD corrections will be discussed with the
sum rules in section III, since they play a major role in the phenomenology of
polarized DIS. \\

The rest of this review is structured as follows. In section II, we discuss
theoretical models for the spin-averaged (unpolarized) distributions and the
spin-weighted (polarized) distributions. The important aspects of how the
theoretical models can be compared to data is discussed in detail throughout
the section. The experimental data and comparison of theory and experiment
is covered in section III. Physical consequences of this comparison and the
open physics questions are addressed. Finally, we outline a set of experiments
which can be used to address these questions and further refine our
understanding of the constituents' contributions to the spin of nucleons. \\

\section{THEORETICAL BACKGROUND}

\subsection{Unpolarized Distributions}

\subsubsection{Formalism} 

In the last section, we discussed the unpolarized and polarized structure
functions for parton distributions in the context of the operator product
expansion and their extraction from deep-inelastic scattering (DIS) data.
In this section, we will outline the details of modeling the constituent
distributions and discuss the possible relations between the unpolarized and
polarized versions. The important aspects of factorization and evolution will
be covered, as they are crucial to relating the theory to experimental data. \\

Due to the models of the polarization mechanism, \cite{FJ, CS} and the possible
relation of the polarized to the unpolarized distributions, knowledge of the
unpolarized distributions is useful before one analyzes the polarized case.
Naturally one can extract the $x$-dependent polarized distributions from the
data directly, \cite{BaT, GS} but the former method has a more direct
theoretical connection. Normally, one assumes an SU(6) symmetric wave function
for the proton, from which the polarized valence distributions can be found
from a suitable parametrization of the unpolarized ones, $u_v(x)$ and $d_v(x)$.
Most early models of the polarized sea and gluon distributions also used this
assumption as a starting point. \cite{ChS, QRRS}

For purposes of studying the quark properties of nucleons, it is convenient
to distinguish between the ``valence'' and the ``sea'' quarks of the proton.
The valence quarks carry the main quantum numbers of the nucleon. For the up
and down quark densities in a nucleon, we can write
\begin{eqnarray}
&  &
u(x,Q^2)=u_v(x,Q^2)+u_s(x,Q^2) \nonumber \\
&  &
d(x,Q^2)=d_v(x,Q^2)+d_s(x,Q^2).\label{2.1}
\end{eqnarray}
The valence distributions are the flavor nonsinglet components of the quark
distributions in the proton and are normalized so that
\begin{eqnarray}
&  &
\int_0^1 dx\,u_v(x,Q^2)=2 \nonumber \\
&  &
\int_0^1 dx\,d_v(x,Q^2)=1. \label{2.2}
\end{eqnarray}
The neutron, of course, exchanges the roles of the up and down quarks, since
it consists of the ``udd" combination of valence quarks.
All other flavor components will be included in the sea. In general, these
distributions will be denoted by their flavor quantum numbers; $\bar u(x,Q^2)$,
$\bar d(x,Q^2)$, $u_s(x,Q^2)$, $d_s(x,Q^2)$, $s(x,Q^2)$, $\bar s(x,Q^2)$, etc. 

For unpolarized distributions, there presently exists a large body of
phenomenological knowledge concerning the various parton densities, which
have been compiled into various sets of models, which will be discussed
shortly. The differences between the different parameterizations of the
unpolarized distributions are usually minor compared to the uncertainties in
data from which they are generated. \\

\subsubsection{Evolution}

The $Q^2$ evolution of the unpolarized (spin-weighted) structure functions is
governed by the GLAP equations, \cite{GLAP} which in the leading logarithm
approximation take the form
\begin{eqnarray}
&  &
{d\over dt}\left[Q_v(x,t)\right]=P_{qq}(x)\otimes Q_v(x,t) \\ \nonumber
&  &
{d\over dt}\left[Q(x,t)\right]=P_{qq}(x)\otimes Q(x,t)+2N_f P_{qG}(x)\otimes
\tilde G(x,t) \\ \nonumber
&  &
{d\over dt}\left[\tilde G(x,t)\right]=P_{Gq}(x)\otimes Q(x,t)+ P_{GG}(x)
\otimes \tilde G(x,t), \label{2.3}
\end{eqnarray}
where,
\begin{eqnarray}
&  &
Q_v(x,t)=x\>q_v(x,t), \\ \nonumber
&  &
Q(x,t)=\sum_i x\>q_i(x,t), \\ \label{2.4}
&  &
t={2\over 11-{2\over 3}N_f}\ln\left[\ln(Q^2/\Lambda^2)/\ln(Q^2_0/\Lambda^2)
\right], \nonumber
\end{eqnarray}
and
\begin{equation}
P(x)\otimes Q(x,t)\equiv \int^1_0 {dz\over z}P(z)Q\left({x\over z};t\right).
\label{2.5}
\end{equation}
The probability kernels, $P_{ij}(x)$, for three quark flavors are given by 
\begin{eqnarray}
&   &
P_{qq}(x)=\frac{4}{3}\left[{1+x^2\over 1-x}\right]_+ +2\delta(1-x) \\
&   &
P_{qG}(x)=\frac{1}{2}\bigl[x^2+(1-x)^2\bigr] \\ \label{2.6}
&   &
P_{Gq}(x)=\frac{4}{3}\Biggl[{{1+(1-x)^2}\over x}\Biggr] \\
&   &
P_{GG}(x)=3\left[\left({x\over {1-x}}\right)_+ +{{1-x}\over x}+x(1-x)+\frac
{3}{4}\delta(1-x)\right]. \nonumber
\end{eqnarray}
The $(1-x)_+$ distribution renormalizes the kernels to avoid divergence at
$x=1$ and is defined by:
\begin{equation}
\int^1_0 {{dx\>f(x)}\over {{(1-x)}_+}}\equiv \int^1_0 dx {{f(x)-f(1)}\over
{1-x}}. \label{2.7}
\end{equation}

These equations imply that even if there is no initial gluon distribution 
generated at the nonperturbative level, a positive $G(x)$ will be generated
by quark Bremsstrahlung, since $P_{Gq}(x)$ is everywhere positive. \\

The next-to-leading order (NLO) splitting functions have been calculated for
the unpolarized distributions. \cite{FP} The separation of the non-singlet
(related to the valence distribution) and the singlet evolution provides a
convenient motivation for the separation of the valence and sea quarks in the
analysis. Thus, the evolution equations allow us to determine the $Q^2$
dependence of the valence and sea densities. We can also investigate the
small-$x$ behavior of the valence quark and gluon evolution, which play a 
significant role in models of the nucleon. There seem to be differences in
some of the NLO evolution codes used to generate unpolarized structure
functions at small-$x$, primarily due to truncation errors in the NLO terms,
i.e. in the NNLO terms. \cite{Blum96a} These manifest themselves most
dramatically at low $x$ and high $Q^2$. This is a minor difficulty relative to
other uncertainties of the structure functions at small-$x$, as will be
discussed later. \\

\subsubsection{Existing Distributions From Data}

Unpolarized parton distributions are determined from global analyses of data
for a large range of processes in as wide of a kinematic range as possible.
There have been various groups who have generated these distributions.
\cite{MRS, GRV, Lai} The improvements in these distributions have come from
more precise data in a wider kinematic domain. \cite{HERA} Theoretical progress
in calculating higher order contributions, both logarithmic (to $\alpha_s$)
and power law in $Q^2$ (higher twist), have increased the accuracy to which
these distributions can fit the data. These also help obtain a more rigorous
test of various aspects of QCD. The differences in the unpolarized
distributions are as follows. \\

The ``Martin, Roberts and Stirling" (MRS) models \cite{MRS} extract the
unpolarized distributions from data, normalized to a fixed $Q_0^2$ value,
normally around 4.0 GeV$^2$. The valence and each flavor of sea quarks are
separately parametrized and the excess of anti-down over anti-up quarks is
built in to the analysis. The parton distributions have the general form:
\begin{equation}
xD(x)=Nx^\alpha \>(1-x)^\beta \>(1+ax^{1\over 2}+bx), \label{2.8}
\end{equation}
with the unkonown factors determined by various data. The gluon density is
chosen to be finite at $x=0$ and the sea has a Regge type behavior at this
limit.

The ``Gl\"{u}ck, Reya and Vogt" (GRV) distributions \cite{GRV} derive $x$ and
$Q^2$ dependent distributions from the data. The valence parametrization is
done separate from the sea and the up and down flavors along with the gluons
include a logarithmic dependence at small-$x$. The strange sea is modeled
separately from the lighter flavors. The form for the valence distributions is:
\begin{equation}
xQ(x,Q^2)=N(Q^2)x^{\alpha(Q^2)} \>(1-x)^{\beta(Q^2)}\>\bigl[1+a(Q^2)x^
{1\over 2}+b(Q^2)x\bigr] \label{2.9}
\end{equation}
and the general form for the other unpolarized distributions can be written as:
\begin{equation}
xS(x,Q^2)=N(Q^2)x^{\alpha(Q^2)}\>(1-x)^{\beta(Q^2)}\>\bigl[1+a(Q^2)x+b(Q^2)x^2
\bigr]\Bigl[\ln({1\over x})+f(\ln({1\over x}),Q^2\Bigr]. \label{2.10}
\end{equation}
Most of the parameters are $Q^2$ dependent, evolved both in leading and next-to
leading orders.

The CTEQ distributions \cite{CTEQ}, like the MRS, are normalized to a fixed
$Q_0^2$. They have the general form:
\begin{equation}
xQ(x)=A_0x^{A_1} \>(1-x)^{A_2} \>(1+A_3x^{A_4}), \label{2.11}
\end{equation}
with the parameters determined by fits to data.
The MRS and CTEQ distributions are evolved from the $Q_0^2$ value by using the
NLO GLAP equations. \cite{FP, Blum96a} All of these distributions are useful
for calculating observables characteristic of the unpolarized scattering
experiments. Further, if suitable assumptions are made about the polarized
distributions, then their $x$ dependence can also be extracted similar to
these parametrizations. \\

\subsubsection{Unpolarized Distributions at Small-x}

Differences in the above distributions for the most part lie in the small-$x$
region. Recent data from HERA has shed light on the small-$x$ behavior of the
parton distributions, but questions of the parametrization and evolution of
this behavior remain controversial. \cite{HERA} 
Since the experimental errors in most polarized DIS experiments are the
largest in this region, there is no apparent advantage that one unpolarized
distribution has over the others in terms of generating the corresponding
polarized distributions. Most of these distributions build in a parametrization
for the difference in anti-up and anti-down quarks, first discovered by the
New Muon Collaboration \cite{NMC} measurement of the Gottfried sum rule at
CERN \cite{GSR} and later confirmed by the NA51 measurement of the $pp/nn$
asymmetry in Drell-Yan. \cite{NA51} For further discussion on the asymptotic
behavior of unpolarized distributions at small-$x$, see Webber \cite{Webber}
and Lopez, \etal, \cite{Lopez} The polarized distributions' small-$x$ behavior
will be discussed in the section covering the polarized sea. \\

\subsection{Polarized Parton Distributions}

\subsubsection{Factorization} 

The usual application of the QCD-parton model assumes that the appropriate
distributions are measured in one set of processes. Predictions are
then made for other processes by using factorization. For the purpose of
estimating the impact of possible experiments, it is convenient to have
model distributions which represent an informed guess concerning the allowable
range of distributions.

The two basic ideas which will help us in constructing these models are:
\begin{enumerate}
\item There is a strong connection between the spin of the valence quarks
and the spin of the proton.
\item There are spin-dependent forces between the constituents in the proton
which influence the shape of the spin-weighted distributions.
\end{enumerate}
A successful description of the baryons within the framework of the ``naive'' 
constituent quark model suggests strongly that, in some approximation,
a large part of the spin of the proton is associated with its valence quarks.
\\

When extracting information from hard-scattering processes, such as DIS,
it is necessary to write the measured cross sections as a product of factors
calculable in PQCD and non-perturbative (usually extractable from data) terms.
Such a process is known as factorization. Factorization theorems exist for many
processes including: \cite{Collins}
\begin{itemize}
\vspace*{-2ex}
\item Inclusive DIS
\vspace*{-2ex}
\item Semi-inclusive production of jets, heavy quarks
\vspace*{-2ex}
\item Drell-Yan: lepton pair production from hadron-hadron interactions
\vspace*{-2ex}
\item High $p_T$ processes: jet production; hadron production; direct photons
\vspace*{-2ex}
\item Hadron-hadron to heavy quark inclusive production
\vspace*{-2ex}
\item $e^+e^-$ to jets
\vspace*{-2ex}
\item Elastic processes
\end{itemize}

The importance of factorization is that it separates the short distance
behavior of the cross sections, calculable in perturbative QCD and the long
range non-perturbative physics. The non-perturbative part should, in principle,
be guage-independent, to ensure that it is measurable.
In principle, one can define factorized quantities which are gauge dependent,
but it is desirable to have them gauge-independent, since the uncalculable
quantities must be measurable. In polarized DIS, there are two schemes which
are used to define the quark spin densities: (1) the gauge-invariant and
(2) the chiral-invariant schemes. Since chirality is involved in the 
spin-dependent processes, neither of these is inherently superior over the
other. \cite{Cheng95} A detailed discussion can be found in Cheng's review.
\cite{Cheng96} \\

Gauge-invariant scheme: \cite{BQ} \\

According to an operator product expansion analysis (OPE), (see section I)
there is no gluonic operator contributing in leading order to $g_1^p$.
The hard-gluonic contributions and the quark spin densities are $k_{\perp}$
factorization dependent. The factorization scheme which respects the OPE
is such that the hard gluons do not contribute to the quark spin densities,
but the hard gluon component of spin does perturbatively generate a negative sea
polarization due to the axial anomaly in the triangle graph for $j_5^{\mu}$
between external gluons. Thus, the anomaly term becomes part of the 
photon-gluon cross section and we would expect to see a correspondingly large
negatively polarized sea, due to the gluon Bremsstrahlung which creates it.
Chiral symmetry is broken in this scheme. \\

Chiral-invariant scheme: \cite{CCM} \\

In this scheme, the polarized quark densities are factorized into a gauge
dependent piece and the gluon axial anomaly is explicitly separated out.
Here, the measured quantities extracted from data and sum rules are modified by
the anomaly term. This accounts for the observed differences between the
first moments (integrals) of $F_1$ and $g_1$. The gluon anomaly contributes
directly to the quark spin densities via the term $\Gamma(Q^2)$ added to each
of the quark flavors. Large negativity of the
sea is not required, but a smaller polarized sea could suggest a larger
polarized glue for the naive parton model for the Ellis-Jaffe sum rule to be
valid. Recent data indicate that this is likely not the case. \\

The key differences in the two schemes lie in the $k_\perp$ factorization of
the quark spin-density and the hard photon-gluon cross section. Then, the
factorization prescription is determined by choice of the ultraviolet cutoff.
 In any case, it has been shown that $g_1$ is independent of which
scheme is used. In both schemes, the polarized gluon density plays a role in
the contribution of each quark flavor to nucleon spin. Should $\Delta G$ be
small for the $Q^2$ values of present data, the flavor dependence of spin
contributions will be approximately the same in both schemes. Recent
treatments by Goshtasbpour and Ramsey \cite{GR95, GR96} extract the flavor
dependent information from data using both factorization schemes by setting
$\Delta G=0$ in one of the models considered. These results will be outlined
in a later section.
Thus, the controversy remains as to which of these is more appropriate in
explaining the constituent spin problem. Further experimentation will determine
whether it is the sea or glue that is large enough to explain why the
quark contribution to the nucleon spin is smaller than unity. \\

The choice of a factorization scale, the scale at which the distributions will
be used in the calculation of various inclusive processes, \cite{CelS} is
consistent in all of the experimental and theoretical analyses. \\

\subsubsection{Sum Rules}

After factorization, the next major step in extracting spin parton densities
from the data is by use of the sum rules. In section I, we discussed the
hadronic tensor and its relation to the cross sections of DIS.
The targets in DIS are generally characterized by sets of conserved quantum
numbers. These can often be built into the analysis by forming combinations
of the cross sections which characterize these conserved quantities. In this
way, theory and experiment are directly compared. These combinations are
referred to as sum rules. Those sum rules which involve the strong coupling
naturally will have higher order logarithmic (perturbative) corrections and
since most are valid over a wide kinematic range will also have higher twist
($Q^2$ dependent) corrections. In DIS, the flavor dependent parton spin
information can be extracted from the data, provided certain sum rules are
assumed to be valid. There have been recent reviews on the parton model sum
rules, \cite{PMSR} but those which are most pertinent to the polarized
distributions will be discussed here.

For a proton with its spin aligned in the $+z$ direction (along its momentum),
we can impose the parity invariance of the strong interactions and define the
spin-weighted densities as
\begin{eqnarray}
&   &
\Delta q^i(x,Q^2)=q^i_{+/+}(x,Q^2)-q^i_{-/+}(x,Q^2) \nonumber \\
&  &
                 =q^i_{-/-}(x,Q^2)-q^i_{+/-}(x,Q^2) \nonumber \\
&  &
                 =\eta_i(x,Q^2)q^i(x,Q^2), \label{2.20}
\end{eqnarray}
where, for a given flavor of quark, the factor $\eta_i(x,Q^2)$ is called the
{\it polarization} of that flavor. For an arbitrary direction of spin, the
component of spin along the $z$ (momentum) direction is the helicity. The
spin-averaged (unpolarized) distributions are just the sums of these helicity
states. If we consider the proton wave function as characterized by momentum
$p_{\mu}$ and spin $s_{\mu}$, the polarized distributions integrated over all
$x$ can be represented in terms of the Dirac matrices $\gamma^{\mu}$ and
$\gamma_5$ by:
\begin{equation}
\langle \Delta q_i s^{\mu}\rangle=\langle ps\mid \bar{q} \gamma^{\mu} \gamma_5
q_i\mid ps\rangle/2m, \label{2.21}
\end{equation}
where $m$ is the mass of the particle. The related axial-vector current
operators, $A_{\mu}^k$, are members of an SU(3)$_f$ octet, whose non-zero
elements provide relations between the polarized distributions and data. The
non-vanishing matrix elements of these operators define measurable
coefficients, $a^k$, which provide some of the sum rule constraints used to
extract parton spin information from the data. These are defined by
\begin{equation}
\langle ps\mid A_{\mu}^k\mid ps\rangle=s_{\mu}a^k \label{2.22}
\end{equation}
where the $a^k$ are non-zero for $k=$0, 3 and 8. The matrix elements, $a_k$ are
determined from weak decays of processes where flavor changes occur. 

The $a_3$ matrix element, measurable in nucleon beta decay, occurs in the
Bjorken Sum Rule (BSR), which is based on isospin invariance. It is considered
to be fundamental to QCD and its validity has been the basis for much of the
recent QCD analysis on polarized DIS. \cite{EK96, CR, GR95} The polarized
version of the BSR can be written as:
\begin{equation}
I^p-I^n\equiv \int_0^1 dx\> (g_1^p-g_1^n)={{a_3}\over 6}(1+\alpha_s^{corr}),
\label{2.23}
\end{equation}
where $\alpha_s^{corr}$ are higher order logarithmic corrections, calculable in
QCD. To $O(\alpha_s^4)$ these are:
\begin{equation}
\alpha_s^{corr}\approx ({{\alpha_s}\over {\pi}})+3.5833({{\alpha_s}\over
{\pi}})^2+20.2153({{\alpha_s}\over {\pi}})^3+130({{\alpha_s}\over {\pi}})^4,
\label{2.24}
\end{equation}
where the last term is estimated. \cite{Larin} Ellis, \etal, have used the
method of Pad\'{e} approximates to estimate the higher order corrections to
the BSR. \cite{Ellis} Their result is consistent with the term quoted above.

The matrix element $a_8$ is determined by the weak decay constants, $F$ and
$D$, which are constrained by hyperon decay experiments. In terms of the
flavor dependent polarized distributions, this can be written as:
\begin{equation}
a_8=\int_0^1 dx\> \bigl(\Delta u+\Delta \bar{u}+\Delta d+\Delta \bar{d}
-2\Delta s-2\Delta \bar{s}\bigr)=3F-D. \label{2.25}
\end{equation}
As Lipkin has pointed out, \cite{Lipkin} since a suitable hyperon model does
not yet exist, one must be careful about imposing the SU(3)$_f$ symmetry
and this constraint on the polarized sea. Some analyses of the recent DIS
data do not rely heavily on this constraint, but use it to narrow the relative
size of the flavor dependent polarized sea distributions. \cite{GR95}

The current $A_8$ is determined by hyperon decay and its eigenvalue $a_8$ is
related to the polarized distributions by:
$a_8=\langle\bigl[\Delta u_{total}+\Delta d_{total}-2\Delta s_{total}\bigr]
\rangle \approx 0.58\pm 0.02.$
Finally, $a_0$ is related to the total spin carried by the quarks in the proton.
It provides one part of the spin-${1\over 2}$ sum rule, known as the
$J_z={1\over 2}$ sum rule. Thus, in terms of the parton distributions,
\begin{equation}
a_0\approx \int_0^1 dx\> \bigl(\Delta u+\Delta \bar{u}+\Delta d+\Delta \bar{d}
+\Delta s+\Delta \bar{s}\bigr)\equiv \Delta q_{tot}. \label{2.26}
\end{equation}
The approximation above is due to the gluon anomaly, to be discussed later.
Then, the axial currents are related to the structure function $g_1^p$ in the
anomaly-independent form:
\begin{equation}
a_0= 9(1-\alpha_s^{corr})^{-1}\int_0^1 g_1^p(x)\>dx-{1\over 4}a_8-
{3\over 4} a_3\approx \langle \Delta q_{tot}\rangle. \label{2.27}
\end{equation}
The polarized distributions are related to the orbital angular momentum of the
constituents by the $J_z={1\over 2}$ sum rule.
If we define $\Delta G\equiv G_+\>-\>G_-$ and the total orbital angular
momentum of the nucleon constituents about the $z-$axis as $L_z$, then the
$J_z={1\over 2}$ sum rule is:
\begin{equation}
J_z\equiv {1\over 2}\Delta q_{tot}+\Delta G+L_z={1\over 2}. \label{2.28}
\end{equation}
This represents the decomposition of the constituent spins along with their
relative angular momentum, $L_z$. \\

The Ellis-Jaffe sum rule \cite{EJSR} is less fundamental than the
BSR and is based upon the assumption that the strange sea is unpolarized. This
assumption was based primarily on the OZI sum rule, which assumes unitary
symmetry for mesons and baryons. \cite{OZI} It implies that the singlet and
octet contributions to the axial-vector currents are equal, i.e., that
$a_0=a_8$. The OZI sum rule has been shown to be valid for mesons, but data
imply that it fails for baryons. \cite{Ji} Ellis and Jaffe applied this rule
to polarized structure functions to predict their averaged values:  
\begin{equation}
\int_0^1 dx\>g_1^{p(n)}(x)={{a_3}\over {12}}\Bigl[\pm 1+{5\over 3}{{(3F/D)-1}
\over {(F/D)+1}}\Bigr]. \label{2.29}
\end{equation}
The $+$ and $-$ signs refer to the proton and neutron cases, respectively.
$F$ and $D$ are the weak decay constants discussed above. The Ellis-Jaffe
sum rule (EJSR) has been widely discussed, since the EMC DIS data were
published, \cite{EMC} which indicated a severe violation of this sum
rule. As previously mentioned, this created the ``spin-crisis" which has led to
a re-assessment of the early models of proton spin. Violation of the EJSR
(or OZI sum rule assumption for baryons) reduces to the question of how
polarized the strange sea is in the nucleons. \\

Regarding this question, it seems clear from recent data that the strange sea
is polarized opposite to that of the valence quarks, but the size of that
polarization is still under question. There is a theoretical constraint
on the polarization of the strange sea, based on the positivity of the
probability interpretation of the leading order parton distributions in the
naive parton model. This implies an upper limit to the polarized strange
sea contribution, which may affect its contribution to the $J_z={1\over 2}$
sum rule. \cite{Pos} The essence of the positivity constraint is
that the spin carried by the strange sea is bounded by its momentum. There may
be non-perturbative contributions to this limit, which may affect its precise
value. This was proposed by Ioffe and Karliner \cite{IK} to explain violations
of the OZI sum rule for baryons. Gehrmann and Stirling \cite{GS96} have also
pointed out that this limit is only valid to leading order, since at higher 
order, the parton distributions no longer have a strict probabilistic
interpretation. Most of the existing DIS data appear to violate the limits of 
this bound to some extent. This issue is still under question. \\

A recent analysis by Ma \cite{Ma} attempts to resolve the EJSR violation by 
introducing a Wigner rotation term as a scale parameter to reconcile the data
with the naive quark model. This is an interesting idea, but as yet there is
no apparent way to strictly test this assumption against other models with
further experimentation. This may be an area for future exploration. \\

Another possible mechanism which has been proposed to explain the violation
of the EJSR is the instanton contribution to the spin-flip mechanism. This
is a non-perturbative vacuum fluctuation of the gluon field, which gives
rise to an effective screening of the valence quark polarization by the sea.
\cite{DK} Predictions for $g_1^n$ and the Drell-Yan asymmetry are given, so
that there exist experimental tests for this model. \\

There are other sum rules which apply to the transversely polarized structure
function, $g_2$. The naive parton model predicts that $g_2(x,Q^2)$ vanishes
everywhere. \cite{PMSR} The Burkhardt-Cottingham sum rule \cite{BC} is somewhat
less stringent, namely that
\begin{equation}
\int_0^1 dx\>g_2(x)=0. \label{2.30}
\end{equation}
This does not imply that $g_2$ is identically zero, in fact recent data
indicate otherwise.\cite{E154, SMC}

The Wandzura-Wilczek sum rule \cite{WW} relates the structure functions $g_1$
and $g_2$:
\begin{equation}
g_1(x)+g_2(x)=\int_x^1 {{dy}\over y}\>g_1(y). \label{2.31}
\end{equation}
This was obtained by an OPE anaysis of quarks in the infinite momentum frame.
The implication of this sum rule is that the transverse spin of the proton is
carried mostly by the sea quarks at small-$x$. \\

An alternate sum rule for $g_2$ is derived from a field-theoretical framework
and is not dependent upon the OPE, where the hadronic matrix elements are local
operators. This ``ELT sum rule" \cite{ELT} relates the valence components of
$g_{1,2}$ and is exact for each flavor. It has the form:
\begin{equation}
\int_0^1 dx\>x\>[g_1^V(x)+2g_2^V(x)]=0. \label{2.31b}
\end{equation}
There are experimental tests which would give information about this sum rule,
namely, semi-inclusive meson production or jet production from unpolarized DIS.
 \\

\subsubsection{$Q^2$ Evolution}

The $Q^2$ evolution of the polarized structure functions is analogous to that
of the unpolarized functions. The polarized version of the GLAP equations
in the leading logarithm approximation, takes the form
\begin{eqnarray}
&   &
{d\over dt}\left[\Delta Q_v(x,t)\right]=\Delta P_{qq}(x)\otimes\Delta
Q_v(x,t) \\ \nonumber
&   &
{d\over dt}\left[\Delta Q(x,t)\right]=\Delta P_{qq}(x)\otimes\Delta
Q(x,t)+2N_f\Delta P_{qG}(x)\otimes\Delta\tilde G(x,t) \\ \nonumber
&   &
{d\over dt}\left[\Delta \tilde G(x,t)\right]=\Delta P_{Gq}(x)\otimes\Delta
Q(x,t)+\Delta P_{GG}(x)\otimes\Delta\tilde G(x,t), \label{2.32}
\end{eqnarray}
where, similar to the spin-averaged (unpolarized) case:
\begin{eqnarray}
&  &
\Delta Q_v(x,t)=x\Delta q_v(x,t)\quad {\textstyle {and}} \\ \nonumber
&  &
\Delta Q(x,t)=\sum_i x\Delta q_i(x,t). \\ \label{2.33}
\end{eqnarray}
The variable $t$ and the convolution operation have the same form as in the
unpolarized case. The probability kernels, $\Delta P_{ij}(x)$, are given by 
\begin{eqnarray}
&   &
\Delta P_{qq}(x)= {4\over 3}\left[{1+x^2\over 1-x}\right]_+ \\ \nonumber
&   &
\Delta P_{qG}(x)= {1\over 2}(2x-1) \\ \label{2.34}
&   &
\Delta P_{Gq}(x)= {4\over 3}(2-x) \\ \nonumber
&   &
\Delta P_{GG}(x)= 3\left[\left({1+x^4\over 1-x}\right)_+ +
	(3-3x+x^2+x^3)-{7\over 12}\delta(1-x)\right]. \nonumber
\end{eqnarray}
These equations imply that even if there is no initial gluon polarization 
generated at the nonperturbative level, Bremsstrahlung from the valence quarks
will generate a positive $\Delta G(x)$, since $\Delta P_{Gq}(x)$ is everywhere
positive.  Furthermore this mechanism will also polarize the sea.  However,
since the first moments of $\Delta P_{qq}$ and $\Delta P_{qG}$ vanish in the
leading order, $\langle \Delta Q(x,t)\rangle\equiv\int^1_0 dx\,\Delta Q(x,t)$
is constant in $t$ (and hence, $Q^2$). \\

The NLO splitting functions have been calculated and are found to be
renormalization scheme dependent. \cite{NLOE}
Once a renormalization scheme is chosen, these evolution equations determine
the $Q^2$ dependence of the valence, sea and gluon spin densities, which
effectively compare the spin carried by constituents to their momentum. 
These become crucial in comparing the different experimental data and their
consequences at different values of $Q^2$. Bl\"{u}mlein and Vogt have carried
out the NLO evolution of $g_1^{p,n}$ in the $\overline{MS}$ scheme. \cite{BV}
The NLO evolution plays an important role in the small-$x$ behavior of these
structure functions. \cite{WM} This will be discussed later. \\
                                                           
The $Q^2$ evolution of $g_2$ has been investigated. \cite{G2, Tspin} Since
these functions are small and not directly pertinent to the constituent spins,
this will not be discussed in further detail here. Instead, we will proceed
directly to the various polarized constituent distributions. \\

\subsubsection{Valence Quark Models}
           
Fundamentally, we assume that the nucleons are comprised of valence quarks,
whose polarized and integrated distributions are defined by:
\begin{eqnarray}
&   &
\Delta q_v (x,Q^2) \equiv q_v^+(x,Q^2)-q_v^-(x,Q^2) \nonumber \\
&   &
\langle \Delta q_v (Q^2)\rangle\equiv \int_0^1 \Delta q_v(x,Q^2)\>dx,
\label{2.35}
\end{eqnarray}                                        
where $+(-)$ indicates the quark spin aligned (anti-aligned) with the nucleon
spin. In order to construct the polarized quark distributions from the
unpolarized ones, we can start with a modified 3-quark model based on an SU(6)
wave function for the proton. This model is based on flavor symmetry of the
u- and d-sea and constructs the valence distributions to satisfy the Bjorken
sum rule. The valence quark distributions can be written in the form:
\begin{eqnarray}
&  &
\Delta u_v (x,Q^2)=\cos \theta_D [u_v(x,Q^2)-{2\over 3}d_v(x,Q^2)], \nonumber
\\
&  &
\Delta d_v (x,Q^2)=-{1\over 3}\cos \theta_D d_v(x,Q^2), \label{2.36}
\end{eqnarray}
where $\cos \theta_D$ is a "spin dilution" factor which vanishes as $x\to 0$
and becomes unity as $x\to 1$, characterizing the valence quark helicity
contribution to the proton. \cite{QRRS, CK} Normally, the spin dilution factor
is adjusted to satisfy the Bjorken sum rule and to agree with the
deep-inelastic data at large $x$.

Two-body spin-dependent forces have a direct influence on the spin-weighted
quark and gluon distributions, and with simple assumptions about their
parametrizations, Qiu \etal, have derived a form for the valence spin-dilution
factor $\cos\theta_D$. This spin dilution factor has the form
\begin{equation}
\cos\theta_D(x) = \frac{\Delta q_v(x)}{q_v(x)}\approx \bigl[1+N(Q^2)
\cdot x\cdot G(x,Q^2)\bigr]^{-1} \label{2.37}
\end{equation}
where it is assumed that $xq^0_v(x)<<xG^0(x)$ at small values of $x$. The
$N(Q^2)$ factor is a normalization term, which is adjusted so that the valence
distributions satisfy the BSR. Note that at small-$x$, the spin dilution factor
has the form:
\begin{equation}
\cos\theta_D=\Bigl[1+N\cdot A_g\cdot x^{1-\alpha_g}\Bigr]^{-1}\simeq
{1\over {N\cdot A_g}}\cdot x^{\alpha_g-1}, \label{2.38}
\end{equation}
where $A_g$ and $\alpha_g$ are the normalization and small $x$ power
coefficients for the unpolarized gluon distribution, respectively.

The integrated polarized structure function, $I^{p(n)}\equiv
\int_0^1 g_1^{p(n)}(x)\>dx$, is related to the polarized quark distributions by
$$
I^{p(n)}=\frac{1}{18}(1-\alpha_s^{corr})\langle\bigl[4(1)\Delta u_{total}
+1(4)\Delta d_{total}+\Delta s_{total}\bigr]\rangle, \label{2.39}
$$
where the QCD corrections ($\alpha_s^{corr}$) are given in Eqn. (41).
In terms of the polarized distributions and the assumptions of a flavor
symmetric polarized $u$ and $d$ sea, the BSR can be reduced to:
\begin{equation}
\int_0^1 [\Delta u_v(x,Q^2)-\Delta d_v(x,Q^2)]\>dx=a_3(1-{{\alpha_s}\over
\pi}+h.o.c.). \label{2.41}
\end{equation}                                         
Thus, the valence contributions can be determined uniquely by this model. 
Any of the unpolarized distributions in principle can be used to generate the
valence quark distributions, evolved to the $Q^2$ scales of each
experiment. These all agree for $x\geq 0.05$, but have subtle differences for
the smaller $x$ values. The spin dilution factor is determined from the BSR,
and differences in the unpolarized distributions are compensated by adjusting
the normalization factor $N$ in the spin dilution term. Thus, the valence
distributions are not sensitive to the unpolarized distributions
used to generate them. The consistency of the resulting polarized distributions
can be checked by comparing them with the value generated for the ratio of
proton and neutron magnetic moments:
\begin{equation}
{{\mu_p}\over {\mu_n}}={{2\langle \Delta u_v\rangle-\langle \Delta d_v\rangle}
\over {2\langle \Delta d_v\rangle-\langle \Delta u_v\rangle}}\approx
-{3\over 2}. \label{2.42}
\end{equation}
Using the values $\langle \Delta u_v\rangle=1.00\pm 0.01$ and $\langle \Delta
d_v\rangle=-.26\pm 0.01$, both the BSR and magnetic moment ratio are satisfied.
This also yields a spin contribution from the valence quarks equal to $0.74\pm
0.02$, consistent with other treatments of the spin content of quarks.
\cite{lattice, Li} The quoted errors arise from data errors on $g_A/g_V$
and any small differences remaining in the choice of the unpolarized
distributions used to generate the $\Delta q_v$ terms. The original analysis
by Qiu, \etal, \cite{QRRS} effectively reached the same conclusion. \\

By an appropriate assumption regarding the relation between the polarized and
unpolarized distributions, an $x$-dependent set of polariaed valence
distributions can be generated. Ma \cite{Ma} has used a quark-spectator
theoretical model to generate an alternate set of polarized valence
distributions. These include Wigner rotation parameters which are fit to the
data. Thus, the approach for generating the valence terms is different than the
above model and comparison would be parameter dependent. \\

\subsection{Models of the Polarized Gluons}

\subsubsection{Overview of Models}

The gluons are polarized through Bremsstrahlung from the quarks. The integrated
polarized gluon distribution is written as
\begin{equation}
\langle\Delta G\rangle=\int_0^1 \Delta G(x,Q^2)\>dx=\int_0^1
[G^+(x,Q^2)-G^-(x,Q^2)]\>dx, \label{2.43}
\end{equation}
where the $+(-)$ indicates spin aligned (anti-aligned) with the nucleon, as
in the quark distributions. We cannot determine {\it a priori} the size of the
polarized gluon distribution in a proton at a given $Q^2$ value. The evolution
equations for the polarized distributions, indicate that the polarized gluon
distribution increases with $Q^2$ and that its evolution is directly related to
the behavior of the orbital angular momentum, since the polarized quark
distributions do not evolve in $Q^2$ in leading order. \cite{Ramsey89}
Thus, one can assume a particular form for the polarized gluon distribution at
a given $Q_0^2$ and evolve it to the higher $Q^2$ values of the data.
Until we can experimentally check its consistency with data which are
sensitive to $\Delta G(x,Q^2)$ over a particular $Q^2$ range, we must assume
models for $\Delta G$ to analyze the spin properties of parton distributions.
Initial analyses of the EMC data led to speculation that the integrated gluon
distribution may be quite large, even at the relatively small value of
$Q^2=10.7$ GeV$^2$. \\

Recent data from the E704 group at Fermilab \cite{E704} indicate that the
polarized gluon distribution is likely not very large at the $Q^2$ values of
present data. With this in mind, there are two feasible models for a small
$\Delta G$, namely:
\begin{eqnarray}
&  &
(1) \qquad \Delta G(x)= x\> G(x), \nonumber \\
&  &
(2) \qquad \Delta G(x)=0. \label{2.44}
\end{eqnarray}
The first implies that the spin carried by gluon is the same as its momentum,
motivated by both simple PQCD constraints and the form of the splitting
functions for the polarized evolution equations. The second provides
an extreme value for determining limits on the values of the polarized sea
distribution, assuming a positively polarized gluon distribution. \\

There are models of $\Delta G$ which result in at least some negativity to
the polarized gluons. Jaffe's model \cite{Jaffe96} is based on a constituent
quark picture, where interactions with gluons are considered. This leads to a
sizable negative polarization of the gluons and is naturally quite speculative. 
The model of Kochelev \cite{Kochelev} is a non-perturbative model, which 
analyses vacuum fluctuations in the gluon field, called ``instantons". The
kinematic analysis results in a polarized gluon distribution which has a
negative component at small-$x$ and a small positive component at larger $x$.
The total integrated distribution is slightly negative, however. Both of these
negative $\Delta G$ models would further compound the ``spin crisis" problem
if the chiral invariant factorization scheme is used. \\

The various gluon scenarios have been summarized by Di Salvo. \cite{DiS}
Nowak, \etal, have done an analysis on the data, including the axial anomaly,
along with instanton-based q-q- interactions. \cite{Nowak}
They conclude that present data imply a positive $\Delta G$ and that the
structure function $g_1^n$ is highly sensitive to the sign of the polarized
gluon distribution. This provides a possible test of these models. This is
discussed in more detail in section III. \\

\subsubsection{Gluon Anomaly}

As we discussed in the factorization section, if a chiral-invariant
factorization scheme is used, the polarized gluon distribution has an effect on
the quark spin distributions via the axial anomaly, to be highlighted here.
Using the same helicity notation as with the quarks and imposing
parity invariance of the strong interactions we write
\begin{eqnarray}
\Delta G(x,Q^2)=G_{+/+}(x,Q^2)-G_{-/+}(x,Q^2) \\
=G_{-/-}(x,Q^2)-G_{+/-}(x,Q^2), \label{2.45}
\end{eqnarray}
which appears in the calculation of spin-related observables involving polarized
protons in the same manner as the $\Delta q_i(x,Q^2)$. The model of $\Delta G$ 
that is used has a direct effect on the measured value of the quark
distributions through the gluon axial anomaly. \cite{Anom, CCM} In QCD, the
U(1) axial current matrix element $A_{\mu}^0$ is not strictly conserved,
even with massless quarks. Hence, at two loop order, the triangle diagram
between two gluons generates a $Q^2$ dependent gluonic contribution to the
measured polarized quark distributions. This term has the general form:
\begin{equation}
\Gamma (Q^2)={{N_f\alpha_s(Q^2)}\over {2\pi}}\int_0^1 \Delta G(x,Q^2)\>dx,
\label{3.13}
\end{equation}
where $N_f$ is the number of quark flavors. Thus, for each flavor of quark
appearing in the distributions, the measured polarization distribution is
modified by a factor: $\langle\Delta q_i\rangle-\Gamma(Q^2)/N_f$.
The quark spin contributions the depend indirectly on $\Delta G$ if the chiral
invariant factorization scheme is used.
In order for us to determine the quark contributions to the spin of the
nucleons, it is necessary for us to know the relative size of the polarized
gluon distribution. If we base our analysis solely on the naive quark model,
then $\sum \Delta q \to 1$ and $\Delta G$ may be quite large to be
consistent with EMC data. If we consider the polarized distributions of Qiu
\etal, \cite{QRRS} a reasonably sized $\Delta G$ is possible if the sea has a 
suitably negative polarization. In section III, we will consider two possible
models for calculating the anomaly contribution: (1) $\Delta G=xG$ (indicating
that the spin carried by gluon is equal to its momentum) and (2) $\Delta G=0$,
which is chosen to estimate a bound the distributions. Present data seem to
imply that anomaly effects, and thus the overall integrated polarized gluon
distribution, is limited at these energies. \cite{E704} \\

In the first moments of the parton densities, the contribution of polarized
gluons to the integral of $g_1^p$ is to produce an effective density
\begin{equation}
\langle \Delta q_i\rangle_{exp}=\langle \Delta q_i\rangle-{{\alpha_s
(Q^2)}\over {4\pi}}\langle \Delta G(Q^2)\rangle \label{2.15}
\end{equation}
for each flavor in the sea. The gluons change the measured net spin
of the sea quarks and antiquarks by an amount
\begin{equation}
\sum_i(\langle \Delta q_i \rangle_{exp}-\langle \Delta q_i\rangle )=
{{N_f \alpha_s(Q^2)}\over {2\pi}}\langle \Delta G(Q^2)\rangle \label{2.16}
\end{equation}
which may or may not be large.
For a more detailed discussion of the anomaly term, see Cheng. \cite{Cheng96} \\

\subsubsection{Role of Orbital Angular Momentum in Nucleon Spin Content}

It is clear from the $J_z=\frac{1}{2}$ sum rule that the angular momentum
accounts for the amount of nucleon spin which is not carried by either the
quarks or gluons. However, the orbital motion may play an even more important
role than this implies. \cite{Berlin} The extended nature of the relativistic
proton and its orbital motion may be responsible for the single spin
asymmetries seen in the data \cite{E704} for inclusive pion production.
Furthermore, there are striking similarities between the inclusive hyperon
polarization from unpolarized $pp$ scattering and the orbital effects which
explain the single spin asymmetries. This is a topic for further study and is
being carried out by the Berlin group (Boros, \etal). Troshin and Tyurin
\cite{TT} have mentioned that single spin asymmetries, which could be measured
at the Relativistic Heavy Ion Collider (RHIC) at Brookhaven, would provide a
good measure of the relative size of the orbital motion of the constituents. \\
               
Ji, \etal, \cite{Lz} have derived evolution equations for the quark and gluon
orbital angular momenta and have concluded that the asymptotic value for
fractions of spin carried by quarks and gluons are $3n_f/(16+3n_f)$ and
$16/(16+3n_f)$ as $Q^2\to \infty$. Ji \cite{Ji} has also derived a sum rule for
the orbital angular momenta and proposes a possible way of measuring these
quantities. (See also, Radyushkin \cite{Rad}). Thus, the orbital motion may be
more interesting than originally thought. Naturally, if future experiments
allow precise measurement of the spins carried by quarks and gluons, the
$J_z=\frac{1}{2}$ sum rule will provide another test of these angular momentum
models. \\

\subsection {Models of the Polarized Sea}

\subsubsection{Overview of Models} 
      
There have been many approaches to extraction of the polarized sea from DIS
data. \cite{CR, EK95, GR95, BT} Most of these are similar in the
global analysis, where the goal is to find the fraction of nucleon spin carried
by each flavor of the sea. Extraction of the $x$-dependent distributions
generally follows along different lines, depending on the treatment of the 
polarized glue and the small-$x$ behavior. \\

The polarization of the sea occurs by gluons that are emitted by gluon
Bremsstrahlung and by quark-antiquark pair creation. The total sea for three
flavors is merely the sum of contributions from the flavors. This is written
for the spin averaged and spin weighted cases as
where
\begin{eqnarray}
S(x)=u_s(x)+\overline u(x)+d_s(x)+\overline d(x)+s(x)+\overline s(x) \\
\Delta S(x)=\Delta u_s(x)+\Delta \overline u(x)+\Delta d_s(x)+\Delta 
\overline d(x)+\Delta s(x)+\Delta \overline s(x), \label{3.5}
\end{eqnarray}
where the valence quark contributions are omitted. The data give information
about the integrated distributions which in the polarized case are
\begin{equation}
\langle\Delta S(Q^2)\rangle\equiv\langle[\Delta u_s(Q^2)+\Delta\bar{u}(Q^2)+
\Delta d_s(Q^2)+\Delta \bar{d}(Q^2)+\Delta s(Q^2)+\Delta \bar{s}(Q^2)]\rangle.
\label{3.2}
\end{equation}

For the unpolarized structure functions, the strange quarks in the sea are
often treated separately \cite{GRV} to account for the excess
of $\bar {d}$ over $\bar {u}$, which violates the Gottfried sum rule. In the
polarized case, the heavier mass of the strange quarks will likely make them
harder to polarize. In fact, the recent analyses of the DIS data agree that
$\Delta s$ is smaller than the lighter flavors, regardless of which approach
is used to model the distributions. The details of extracting the integrated
spin-weighted distributions from the data will be given in the next section.
Here we will outline two possible ways of parametrizing the polarized sea quark
distributions. \\

In the approach of Goshtasbpour and Ramsey \cite{GR96}, is is assumed that the
power of $(1-x)$ is the same for both the polarized and unpolarized sea,
indicating the same large $x$ asymptotic behavior. For each flavor they
assume
\begin{equation}
\Delta q_i(x)\equiv \eta_i xq_i(x), \label{3.6}
\end{equation}
where the polarization terms, $\eta_i$, are a set of flavor dependent
parameters, to be determined by data. Then, knowledge of the unpolarized
distributions will yield the $x$-dependent polarized densities. Then, the
appropriate spin observables can be calculated to compare with data. In the
Gehrmann/Stirling \cite{GS} and Bartelski/Tatur \cite{BaT} approaches, the
polarized distributions are written as:
\begin{eqnarray}
S(x)=A_{us} x^{\alpha_{us}}(1-x)^{\beta_{us}}(1+ax^{1\over 2}+bx) \\
\Delta S(x)=A_s x^{\alpha_{sea}}(1-x)^{\beta_{sea}}(1+ax^{1\over 2}+bx),
\label{3.4}
\end{eqnarray}
and the parameters are extracted from data. \\

In either of these scenarios, it is found that data imply a negatively
polarized sea. This negative polarization can be generated in two possible
ways: \\ \\
(1) the perturbative mechanism of the gluon axial anomaly, which breaks chiral
symmetry, or \\
(2) the non-perturbative instanton mechanism, which causes quark helicity
flipping to induce the negative sea. \\ \\

Other theoretical arguments implying a negatively polarized sea involve
non-perturbative spin-spin correlations and the large-$N_c$ chiral Lagrangian
motivated by a Skyrme model of the proton. \cite{Li} \\

A detailed discussion of the sea distributions extracted from data is in the
next major section. \\

\subsubsection{Small-$x$ Behavior}

The small-$x$ behavior of the polarized distributions is crucial to both
understanding the role of perturbative QCD in DIS and extracting the flavor
dependent polarization densities from the data. Recent HERA data for the
unpolarized distributions indicates growth of $F_2$ at small-$x$. \cite{HERA}
Since the polarized distributions are extracted from $F_2$, this has a direct
affect on the polarization of the proton in this kinematic region. (For a
treatment of both unpolarized and polarized structure functions at small-$x$,
see Webber\cite{Webber}) Historically, Regge theory predicted that
$g_1^{p,n}\simeq x^{-\alpha}$, where the axial-vector meson trajectory,
$\alpha$ was in the range $-0.5\le \alpha \le 0.0$. Data indicate that this
prediction is likely valid for $Q^2\le 1$ GeV$^2$, but that perturbative
effects are more appropriate for higher $Q^2$.

There are two popular approaches to analyzing the small-$x$ behavior of the
polarized structure functions. The first is a standard resummation of the
GLAP $Q^2$ logarithms: $\sum_{n,m} \alpha_s(t)^n(\ln Q^2)^m$, implying
\cite{BFR}
\begin{equation}
\Bigl(\ln\mid {1\over x}\mid\Bigr)^p\ll g_1^{p,n}\ll x^{-q}, \label{3.4a}
\end{equation}
for some positive $p$ and $q$. The second is BFKL inspired resummation of
($1/x$) logarithms: \cite{BFKL} \\
 $\sum_{n,m} \alpha_s(t_0)^n(\ln (1/x))^m$ at fixed $Q^2$
where the non-singlet and singlet structure functions behave as: \cite{BER}
\begin{eqnarray}
&  &
g_1^{NS}\sim x^{-0.4}\Bigl({{Q^2}\over {\mu^2}}\mid\Bigr)^{0.2} \\ \nonumber
&  &
g_1^{S}\sim x^{-1.0}\Bigl({{Q^2}\over {\mu^2}}\mid\Bigr)^{0.5}. \label{3.4b}
\end{eqnarray}
The resummation of higher order corrections is important in understanding the
small-$x$ behavior of the polarized structure functions. \cite{BV}
Preliminary data from the SLAC E154 experiment can be fit to a
power law $g_1^n\sim x^{-0.8}$ for $0.02\le x\le 0.1$, but Ratcliffe has
pointed out that the data can be fit equally well to a form:
$g_1^n\sim x^{-0.5}(1-4x)$. \cite{Ratc} This is consistent with an
isospin decomposition of data done by Soffer and Teraev, \cite{ST}
who give a small-$x$ power of $x^{-0.45}$. The small-$x$ extrapolation can
give a net value of the integral $\int_0^1 dx\>g_1^n$ which differs by up to a
factor of two. This makes a large difference in the extraction of spin
information from this data. So far, HERA data has not been able to
differentiate between these scaling models for $F_2$, but future experiments
at HERA or the LHC could reach lower values of $x$ at high enough
$Q^2$. \cite{BH} A recent analysis of $g_1^{p,n}$ via an all-order resumming of
the $O(\alpha_s^{l+1}\ln^{2l}\>x)$ terms in the singlet evolution, indicates
a large uncertainty in the behavior in these structure functions at small-$x$
due to uncalculated correction terms. Thus, it appears that both theoretical
and experimental work must be done in order to isolate the small-$x$ behavior
of the polarized structure functions, and hence, the spin distributions. \\
A crucial problem here is extrapolation of the structure function
$g_1(x)$ to $x\to 0$. The region of small $x$ is particular interesting
since it provides an insight into interface of the perturbative and
nonperturbative regions of QCD. The small-x behavior of structure
function $g_1(x)$  has been described traditionally in the Regge model
by the contribution of $a_1$ trajectory with intercept
$\alpha_{a_1}(0)\simeq 0$.  However, the SMC data point out that
$g_1(x)$ might increase at small $x$. Theoretical background for
the Regge extrapolation is also questionable, since we deal with
the amplitudes with virtual external particles. Indeed perturbative QCD
evolution gives another form for $g_1$ at small $x$, i.e.
$g_1(x)\sim\exp{\sqrt{\ln{1/x}}}$. Other forms of this dependence
are allowed in general Regge analysis with account for
cut--contributions, two--gluon model for Pomeron. Even strongly rising at
$x\rightarrow 0$ dependencies such as $g_1\sim \ln^2{x}/x$ are possible.
In the latter case the integral for the first moment of $g_1$ is divergent.
An important question here is the role of unitarity for the amplitudes with
virtual external particles, i.e. whether it provides any restrictions to the
growth of $g_1$. Thus the problem of extrapolation to $x\to 0$ is important.
The only way to resolve this problem is with the experimental measurements
in the region of $x\sim 10^{-4}-10^{-5}$. Such measurements are possible
with the polarized proton beam at HERA. \\

Another potential contribution to the spin analysis at small-$x$ is the
possible $Q^2$ scaling violation of the asymmetries $A_1^i$ (i=p,n,D). A
detailed analysis done by Gluck, \etal, \cite{GRSV} indicates that this slight
scaling violation could exist at most $x\le 0.25$ for $Q^2\le 4$ GeV$^2$. 
The potential problem here is that the experimental analysis is done using the
assumption that the asymmetries are independent of $Q^2$. It is unclear that
this effect would be significant, since the scaling violation is small and the
values of $g_1^i$ are more sensitive to the small-$x$ fit than the scaling
of $A_1$. This will be discussed in section III. \\

All sets of data are limited in the range of Bjorken $x$ and thus, the
integrals must be extrapolated to $x\to 0.$ Thus, the possibility of existance
of a Regge type singularity at $x\to 0$ is not guaranteed in the analyses.
A significant singularity could raise the value of $g_1^p$ towards the naive
quark model value and could account for some of the discrepancy between the
original EMC data and the Ellis-Jaffe sum rule. In light of the recent HERA
data, there is the possibility that the increase in $F_2$ at small
$x$, even at the lower $Q^2$ values of the E142/E143 data, could indicate a
change in the extrapolated values of these integrals. These possibilities are
a topic for future study. It is also possible that the overall effect of $F_2$
on $g_1^p$ will not alter the integral by any more than the present
experimental errors.
The shape of the polarized gluon distribution at small-$x$ affects the anomaly
term, and thus the overall quark contributions to the integrals. Future
experiments can shed light on the size of this effect, a detail discussed
later. There is still controversy as to whether present data show that anomaly
effects are limited or not, thus leaving open the question of the size of the
polarized gluon distribution at these energies. This will be discussed in
detail in the next section. \\

\section{EXPERIMENTAL CONSEQUENCES}

\subsection{Experimental Overview}

\subsubsection{Recent Data from SLAC, CERN and DESY}

The most recent generation of Deep-Inelastic Scattering (DIS) experiments began
with the European Muon Collaboration (EMC) at CERN in 1987. \cite{EMC} Their
measurement of $g_1^p$ yielded results which were in disagreement with the
Ellis-Jaffe sum rule (EJSR). Naturally, this motivated many theoretical
analyses, as well as plans for further polarized DIS experiments to test the
models which were devised to explain the discrepancy. As a result, experimental
groups at CERN and SLAC performed a number of these experiments. Since 1992,
these groups have succeeded in measuring the corresponding neutron and deuteron
structure functions as well as measuring $g_1^p$ more precisely than the EMC.
In all cases, the statistical and systematic errors were significantly
decreased and a wider range of Bjorken-$x$ values was probed. This improvement
has been promoted by both technological
developments and increased running time.

Beginning in 1993, the SMC group measured $g_1^p$, while the E142 experiment
measured $g_1^n$. This provided a complementary set of measurements to test
both the Bjorken sum rule (BSR), which measures the difference of their
integrals, and the Ellis-Jaffe sum rule, (EJSR), which predicted their separate
integrals. In 1994 and 1995, both the SMC and E143 groups measured $g_1^p$ and
$g_1^d$ to check consistency with former results and provide a ``world" average
for these quantities, so that a more thorough theoretical analysis could be
carried out. A number of theoretical teams performed these analyses. 
\cite{CR, EK95, CL, GR96, BFR} Before the recent E154 and HERMES data were 
released, the neutron data seemed to yield different implications for the
polarized strange sea than the proton and deuteron data. These groups have
measured $g_1^n$ to higher precision and over a wider range of $x$ so that a
revised value was given. This has brought the neutron data closer to the
implications of the other data, but this issue is still not resolved.

A comparison of experimental results and their corresponding theoretical
analyses will be summarized in this section. The physical implications of the
analyses will then be discussed, along with a planned set of future experiments
which will help to resolve some of the unanswered physics questions.

A summary of key measurements in polarized DIS is given in Table I. Note the
complementary measurement of the structure functions as well as the increased
coverage of the $x$ range and decreasing experimental errors. Note that the
E154 \cite{E154} and HERMES \cite{HERMES} data are preliminary.
     
\newpage
\begin{center}
\large
{Table I: Recent DIS Experimental Data} \\
\normalsize

\begin{tabular}{|l||c||r||c||c|}
\hline
 Experiment & Target & $\VEV{Q^2}$ & x-range & $I^{target}$    \\
\hline
  EMC (87)  & p    & 10.7  & $0.1<x<0.7$ & $0.126\pm 0.010\pm 0.015$  \\
\hline
  SMC (93)  & d    &  4.6  & $0.006<x<0.7$ & $0.023\pm 0.020\pm 0.015$ \\
\hline
  SMC (94)  & p    & 10.0  & $0.003<x<0.7$ & $0.136\pm 0.011\pm 0.011$ \\
\hline
  SMC (95)  & d    & 10.0  & $0.003<x<0.7$ & $0.034\pm 0.009\pm 0.006$  \\
\hline
  E142 (93) & n    &  2.0  & $0.03<x<0.6$  & $-0.022\pm 0.007\pm 0.006$ \\
\hline
  E143 (95) & p    &  3.0  & $0.03<x<0.8$  & $0.127\pm 0.004\pm 0.010$  \\
\hline
  E143 (95) & d    &  3.0  & $0.03<x<0.8$  & $0.042\pm 0.003\pm 0.004$  \\
\hline
  E154 (96) & n    &  4.5  & $0.014<x<0.9$ & $-0.037\pm 0.004\pm 0.010$ \\
\hline
  HERMES (96) & n  &  3.0  & ---        & $-0.032\pm 0.013\pm 0.017$  \\
\hline
\end{tabular}
\end{center}

Most of the CERN experiments have used polarized muons, while the SLAC and
DESY (HERMES) experiments have used electrons, being primarily electron
accelerators. As Frois pointed out, \cite{Frois} electron data tend to
be more accurate due to the higher intensities possible, but provide a less
broad kinematic range than muons, due to their smaller energies. The systematic
errors in each are similar. Thus, these experiments provide complementary
information about the polarized structure functions. \\

The targets typically used for these experiments are: ammonia crystals (NH$_3$)
for the proton data, $^3$He for the neutron data and deuterated butanol
($^{15}$ND$_3$) for the deuteron data. More detailed information on targets
is contained in workshop proceedings: D.G. Crabb, N. Horikawa and V.G.
Luppov in the VI Workshop on High Energy Spin Physics, Protvino (Vol. 2), 1996,
and S. Goertz, V.G. Luppov, B. Owen in the 12th International Symposium on High
Energy Spin Physics, Amsterdam, September, 1996. See also the reference of
Crabb and Day. \cite{CD} Szwed \cite{Szwed} has done a partial study on the
effect of the nuclear targets on the reliability and understanding of the data.

Other recent technical developments in polarization experiments include
improvements on Siberian snakes used to retain polarization of accelerating
protons \cite{EIS}, better polarized ion sources and detector improvements.
\cite{TIS} \\

\subsection{Extracting Results from Data}

\subsubsection{Extraction of Polarized Sea and Gluons from Data and
Sum Rules}

Being the earliest analysis, the Close and Roberts approach \cite{CR}
used leading order structure functions and used the earliest data (EMC
proton, E142 neutron and SMC deuteron). They start with the BSR and use
average values for the F and D weak decay constants, discussed in section II.
They introduce an abitrary parameter for the possible higher-twist corrections.
A value for the total spin carried by quarks, $\Delta q_{total}$, was found 
for each experiment. The resulting error bars were quite large, since data
were over a smaller kinematic range and the experimental errors were larger.
Further, the higher order QCD and higher-twist corrections were not known as
well as at present.

Ellis and Karliner \cite{EK95} expanded this analysis by including
higher order QCD and higher-twist corrections and taking a world average of
the data available as of their writing. This included the EMC, E142 and E143
experiments as well as the SMC proton and deuteron experiments. They used the
sum rules to extract flavor dependent information about the proton spin.

Cheng and Li \cite{CL} took a more theoretically motivated approach
by considering a chiral quark model with the Gottfried sum rule violation
built in (asymmetry of anti-down to anti-up quarks). Their experimental inputs
were the value of this asymmetry, measured by the NMC group at CERN \cite{NMC}
and the $\sigma_{\pi N}$ factor from pion-nucleon experiments. Their polarized
sea was completely SU(3) symmetric.  \\

The approach of Goshtasbpour and Ramsey \cite{GR96} included higher order
corrections to the structure functions and did a complete flavor dependent 
analysis, including two different gluon models to investigate the
effects of the gluon anomaly. They also break the SU(3) symmetry of the flavor
dependent sea to explicitly separate the strange sea contribution. This was
repeated for each of the experiments separately and therefore indicated where
the data were consistent and where they disagreed in terms of physical
implications. Being a more explicit analysis, it reflects many of the 
techniques used by the theoretical groups to analyse the data. It will
therefore be outlined shortly as an example. Their 1996 update includes
preliminary data from the E154 and HERMES experimental groups, which will be
covered later.

The Ball, Forte and Ridolfi analysis \cite{BFR} places
emphasis on the low-$x$ behavior of the structure functions and does a moment 
analysis of their evolution. They assume complete SU(3) symmetry and extract 
both quark and gluon spin information from the data. A comparison of the
results from these theoretical approaches will be discussed later.

To give an indication of how the data and sum rules are used to extract the
spin information about the quark and constituents, we will explain the basic
approach of Goshtasbpour and Ramsey (G-R).
The basic source of information for fixing the parameters in the models is the
data on the longitudinal spin-spin asymmetry in deep-inelastic lepton-proton
scattering and the sum rules discussed in section II. The measured integral of
$g_1$ in each experiment constrains the appropriate spin parameters. The
higher-twist corrections \cite{Larin} appear to be negligible at the $Q^2$
values of the data, so they are not included. The additional constraints are
provided by the axial-vector current operators, $a_3$, $a_8$ and $a_0$, as
discussed earlier. The BSR is used to extract an effective $I^p\equiv \int_0^1
g_1^p(x)\>dx$ from all data.

The model of $\Delta G$ that is used has an affect on the quark distributions
through the gluon axial anomaly, which was discussed earlier.
In Table II, two models for $\Delta G$ are considered: (1) $\Delta G=xG$
and (2) $\Delta G=0$. The E154 and HERMES data are preliminary, as reported
in the Amsterdam symposium. \cite{E154, HERMES} \\

The key elements of the G-R approach are:
\begin{itemize}
\vspace*{-2ex}
\item
determine the valence contribution to the spin using the BSR
\vspace*{-2ex}
\item
find sea integrated parton distributions for each flavor by breaking the SU(6)
symmetry with the strange quarks and using the sum rules with data as input
\vspace*{-2ex}
\item
include higher order QCD corrections and the gluon anomaly for each flavor
\vspace*{-2ex}
\item
discuss similarities and differences between the phenomenological implications
of the different experimental results, and
\vspace*{-2ex}
\item
suggest a set of experiments which would distinguish the quark and gluon
contributions to the proton spin.
\end{itemize}

This approach differs from that of others in that the sum rules are used in
conjunction with a single experimental result to extract the spin information
and the flavor symmetric sea is broken while anomaly contributions are
included via the gluon models. The sea breaking parameter, $\epsilon$, is
defined by:
\begin{equation}
\Delta u_{sea}=\Delta \bar{u}=\Delta d_{sea}=\Delta \bar{d}=(1+\epsilon)
\Delta s=(1+\epsilon)\Delta \bar{s}. \label{3.1a}
\end{equation}

The analysis (for each polarized gluon model) proceeds as follows:
\begin{itemize}
\vspace*{-2ex}
\item
Extract a value of $I^p$ from either the data directly or via the BSR in the
form of equation (77),
\vspace*{-2ex}
\item
use Eqn. (44) to extract $a_0$. Then the overall contribution to the
quark spin is found from $\langle \Delta q_{tot}\rangle=A_0+\Gamma$, where
$\Gamma$ is the gluon anomaly term in equation (65),
\vspace*{-2ex}
\item
use the value $a_8$ from the hyperon data to extract $\Delta s$ for the
strange sea,
\vspace*{-2ex}
\item
find the total contribution from the sea from $\langle\Delta q_{tot}
\rangle=\langle\Delta q_v\rangle+\langle\Delta S\rangle$,
\vspace*{-2ex}
\item
determine the SU(3) breaking parameter, $\epsilon$ and the distributions
$\langle \Delta u\rangle_{sea}=\langle \Delta d\rangle_{sea}$ from equation
(76) and the strange sea results and
\vspace*{-2ex}
\item
finally, extract $L_z$ from the $J_z$= 1/2 sum rule.
\end{itemize}

\subsubsection{Comparison of Results from Different Experiments}

Data from SMC \cite{SMC}, SLAC \cite{E143, E154} and DESY \cite{HERMES}
are used to extract information
about the flavor dependence of the sea contributions to nucleon spin.
We can write the integrals of the polarized structure functions,
$I^i\equiv \int_0^1 g_1^i\>dx$ in the terms of the axial-vector currents as:
\begin{eqnarray}
&  &
I^p\equiv \int_0^1 g_1^p(x) dx=\Biggl[{{A_3}\over {12}}+{{A_8}\over {36}}+
{{A_0}\over 9}\Biggr]\Bigl(1-\alpha_s^{corr}\Bigr), \\ \nonumber
&  &
I^n\equiv \int_0^1 g_1^n(x) dx=\Biggl[-{{A_3}\over {12}}+{{A_8}\over {36}}+
{{A_0}\over 9}\Biggr]\Bigl(1-\alpha_s^{corr}\Bigr), \\ \label{3.1}
&  &
I^d\equiv (1-{3\over 2}\omega_D)\int_0^1 g_1^d(x) dx=\Biggl[{{A_8}\over {36}}
+{{A_0}\over 9}\Biggr]\Bigl(1-\alpha_s^{corr}\Bigr)(1-{3\over 2} \omega_D),
\nonumber
\end{eqnarray}
where $\omega_D$ is the probability that the deuteron will be in a D-state.
Using N-N potential calculations, the value of $\omega_D$ is about $0.058$.
\cite{Omega} The BSR can then be used to extract an effective $I^p$ value 
from all data. Comparison of the $I^p_{eff}$ values from each experiment gives
a measure of the validity of the BSR.

Since the evolution splitting functions for the polarized distributions have an
additional factor of $x$ compared to the unpolarized case, early treatments of
the spin distributions assumed a form of: $\Delta q(x)\equiv xq(x)$ for all
flavors. This form of the distributions has been
compared to those extracted from the recent data, using the defined ratio
$\eta \equiv{{\langle \Delta q_{sea}\rangle_{exp}}\over {\langle xq_{sea}
\rangle_{calc}}}$ for each flavor. Any deviation from $\eta=1$ would indicate
that the early models for generating the polarized distributions are inaccurate.
The results are given in Table II.

\newpage
\begin{center} \large
{\bf Table II: Integrated Polarized Distributions: \\
$\Delta G=xG$ (above line), $\Delta G=0$ (below line)}
\end{center}  \normalsize

$$\begin{array}{cccccc}
  Quantity    &SMC(I^p)  &SMC(I^d)  &E154(I^n)  &E143(I^d) &HERMES \cr
              &          &          &           &          &(I^n)  \cr
  <\Delta u>_{sea} &-.077  &-.089   &-.063      &-.068    &-.050 \cr
  <\Delta s>       &-.037  &-.048   &-.020      &-.028    &-.010 \cr
  <\Delta u>_{tot} &0.85   &0.82    &0.87       &0.87     &0.90  \cr
  <\Delta d>_{tot} &-.42   &-.43    &-.39       &-.40     &-.36  \cr
  <\Delta s>_{tot} &-.07   &-.10    &-.04       &-.06     &-.02  \cr
  \eta_u = \eta_d  &-2.4   &-2.8    &-1.9       &-2.1     &-1.5 \cr
    \eta_s      &-2.0    &-3.0      &-1.2       &-1.6     &-0.6 \cr
   \epsilon     &1.09    &0.84      &2.10       &1.41     &4.00 \cr
    \Gamma      &0.06    &0.06      &0.08       &0.08     &0.07 \cr
     I^p        &0.136   &0.129     &0.134      &0.131    &0.135 \cr
  <\Delta q>_{tot} &0.36   &0.29    &0.45       &0.41     &0.52 \cr
  <\Delta G>    &0.46    &0.46      &0.45       &0.44     &0.44 \cr
     L_z        &-.14    &-.11      &-.18       &-.15     &-.22 \cr
 ------& -----& -----& ------& -----& ----- \cr
 <\Delta u>_{tot} & .83     & .80    & .85    & .84   & .88   \cr
 <\Delta d>_{tot} & -.44    & -.45   & -.41   & -.43  & -.39  \cr
 <\Delta s>_{tot} & -.09    & -.12   & -.07   & -.08  & -.04  \cr
     I^p          & .136    & .129   & .134   & .131  & .135  \cr
 <\Delta q>_{tot} & 0.30    & 0.23   & 0.37   & 0.33  & 0.45 \cr
  \Gamma          & 0.00    & 0.00   & 0.00   & 0.00  & 0.00 \cr
     L_z          & 0.35    & 0.39   & 0.32   & 0.35  & 0.28
\end{array}$$

\subsection{Consequences of the Results}

\subsubsection{Physics Consequences}

From these results, it is obvious that the naive quark model is not
sufficient to explain the proton's spin characteristics. Nor is the simple
model for extracting the polarized distributions accurate. Some conclusions
which can be drawn from the data are: \\

(1) The total quark contribution to proton spin is between 1/4 and 1/2. The
errors in generating these results are due mostly to experimental errors and
determination of which model of the polarized gluons to use. \\

(2) The up and down sea contributions seem to agree within a few percent.
However, the proton and deuteron data imply a larger polarized sea with the
strange sea polarized greater than the positivity bound. \cite{Pos}
Interestingly, the SMC proton data are consistent with a recent lattice
QCD calculation of these parameters. \cite{lattice} The results from these
data can be categorized into distinct models, characterized by the size of the 
non-zero polarized sea. \\

(3) The values of $\eta$ deviate considerably from unity for most of the data,
implying that the relation between unpolarized and polarized distributions is
more complex than originally thought. \\

(4) This analysis implies that the anomaly correction is not large.
If the anomaly term were larger, due to a large $\Delta G$, the strange sea
would be positively polarized, while the other flavors are negatively
polarized. There is no known mechanism that would allow this cross polarization
of different flavors. These data imply that $\Delta G$ is of small to moderate
size. Further, even if there are higher twist corrections to the anomaly
at small $Q^2$, the anomaly will not reconcile differences in the flavor
dependence of the polarized sea. \\

(5) The orbital angular momentum extracted from data is also much smaller than
earlier values obtained from EMC data. In fact, a small $\Delta G$ model
implies a correspondingly small orbital angular momentum, although its sign is
still in question. \\

(6) The extracted $I^p$ value is comparable for all data and well within
the experimental uncertainties. This implies agreement about the validity of
the Bjorken Sum Rule. This has been done here by using the BSR to extract an
effective $I^p$, in contrast to other analyses, which use data to extract the
BSR. There is general agreement that the BSR (and thus QCD) is in tact. \\

Clearly, these experiments have contributed to the progress of understanding
the relative contributions of the constituents to the proton spin. They have
probed to smaller $x$ values, while decreasing the statistical and systematic
errors. This, coupled with theoretical progress in calculating higher order
QCD and higher twist corrections have allowed us to narrow the range of these
spin contributions. Although the flavor contributions to the proton spin cannot
be extracted precisely, the range of possibilities has been substantially
decreased (see Table III). The main differences are the questions of the
strange sea spin content and the size of the polarized gluon distribution.
Obviously, more experiments must be performed to determine the relative 
contributions from gluons and various flavors of the sea. \\

\begin{center}
\large                                            
{Table III: Ranges of Constituent Contributions to Proton Spin}
\end{center}
\normalsize

\begin{center}
\begin{tabular}{|c||c||c|}
\hline
  Quantity    &EMC results  &Post-SMC/SLAC \\ \hline\hline
  $\VEV{\Delta u}_{sea}$ &-0.077        &-0.089    \\ \hline
  $\VEV{\Delta s}$       &-0.037        &-0.028    \\ \hline
  $\VEV{\Delta u}_{tot}$ &0.85         &$0.80\to 0.90$ \\ \hline
  $\VEV{\Delta d}_{tot}$ &-0.42        &$-0.35\to -0.45$ \\ \hline
  $\VEV{\Delta s}_{tot}$ &$-0.25\to 0$ &$-0.12\to 0$ \\ \hline
    $I^p$                &0.126        &0.136    \\ \hline
  $\VEV{\Delta q}_{tot}$ &$0\to 1$     &$0.2\to 0.5$ \\ \hline
  $\VEV{\Delta G}$       &$0\to 6$     &$0\to 1.50$   \\ \hline
     $L_z$               &$0\to 6$     &$0\to 1.25$   \\ \hline
\end{tabular}
\end{center}

Table IV shows a comparison of various recent approaches to extraction of the
spin information from data. The key is as follows: BFR \cite{BFR}; CL \cite{CL};
CR \cite{CR}; EK \cite{EK95} and GR. \cite{GR96} \\

\begin{center}                                           
{\bf Table IV. Comparison of Results with Different Models} \\
\end{center}

$$\begin{array}{|cccccc|}
\hline
  Quantity/Model- &BFR(96)  &CL(95)  &CR(93)   &EK(95)   &GR(96) \cr
\hline \hline
  <\Delta u>_{tot} & 0.88   & 0.79   & --      & 0.83    & 0.86  \cr
\hline
  <\Delta d>_{tot} &-0.38   &-0.32   & --      &-0.43    &-0.40  \cr
\hline
  <\Delta s>_{tot} & 0.00   &-0.10   & --      &-0.10    &-0.06  \cr
\hline
     I^p           &0.122   & --     &0.126    &0.133    &0.133  \cr
\hline
  <\Delta q>_{tot} & 0.50   & 0.32   & 0.38    & 0.30    & 0.40  \cr
\hline
  <\Delta G>       & 1.50   & 0.00   & --      & --      & 0.45  \cr
\hline
     L_z           &-1.25   & 0.34   & --      & --      &-0.15  \cr
\hline
\end{array}$$ 

It is clear from these results, that even with varied approaches and
assumptions, the up and down polarized distributions are fairly consistent.
A slightly different approach was taken by Gl\"{u}ck, \etal, \cite{GRSV} in a
next-to-leading order analysis. Their results were comparable to BFR, except
for the polarized strange sea, which agreed with the GR approach.
The range of $\Delta s$ and $\Delta G$ are quite considerable in these models.
Part of the problem is the small-$x$ contributions to the structure functions,
which imply different gluon contributions. Also the lack of knowledge of the
orbital component of motion prevents us from making a statement about the glue
or the strange sea. \\

\subsubsection{$x$-Dependent Distributions}

In order to generate the $x$-dependent distributions, there are two approaches
which were mentioned in section II. Three groups have extracted these
distributions directly from the data. \cite{Alek, BT, GS} All give good
agreement with data, but they differ significantly in the small-$x$ region.
Some are consistent with Regge behavior, while others are not. Goshtasbpour
and Ramsey \cite{GR95, GR96} have used the unpolarized distributions with their
extracted value of $\eta$ and the assumption that: $\Delta q(x)\equiv
\eta xq(x)$ for each of the sea flavors. For the valence distributions,
they have used the model of Qiu, \etal. \cite{QRRS} There is no reason {\it a
priori} to suspect that a global fit to the integrated distributions should 
imply a satisfactory $x$-dependent fit to the data. However, figs. 1 through 3
indicate that this form gives very good $x$-dependent parametrizations for the
polarized distributions, consistent with data and Regge behavior. \\

The differences between the these sets of distributions are at small-$x$,
where the data is most uncertain. It is clear that more DIS experiments should
be performed to probe very small-$x$ to distinguish between models and to
address the controversy regarding which contributions to $g_1$ dominate in 
this kinematic region. It will be required, however, that the error have to be
minimized to distiguish between the various possible powers of small-$x$ for the
polarized structure functions.  \\

\begin{figure}
{\hskip 0.3cm}\hbox{\epsfxsize7.5cm\epsffile{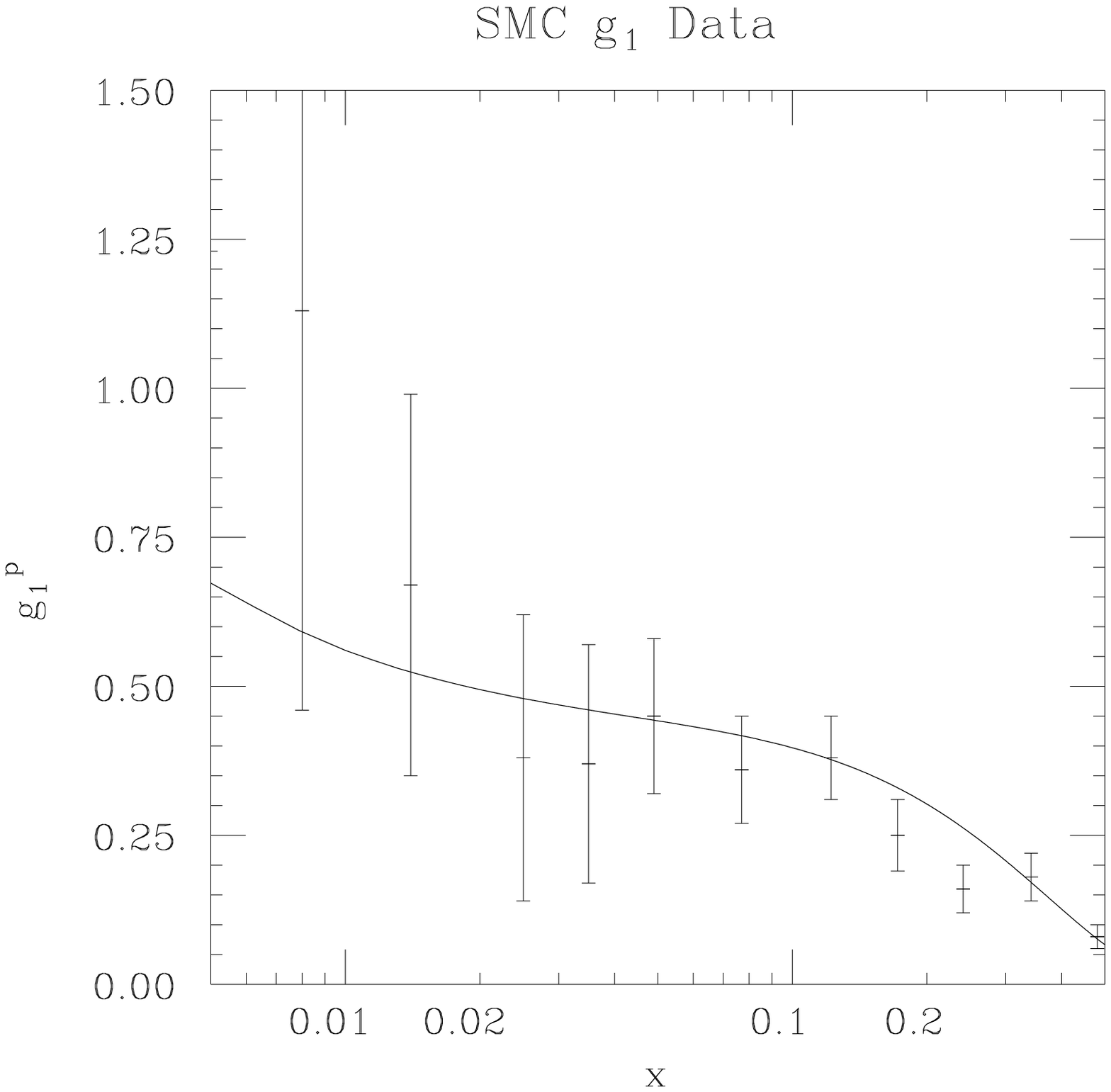}{\hskip 0.3cm}
\epsfxsize7.5cm\epsffile{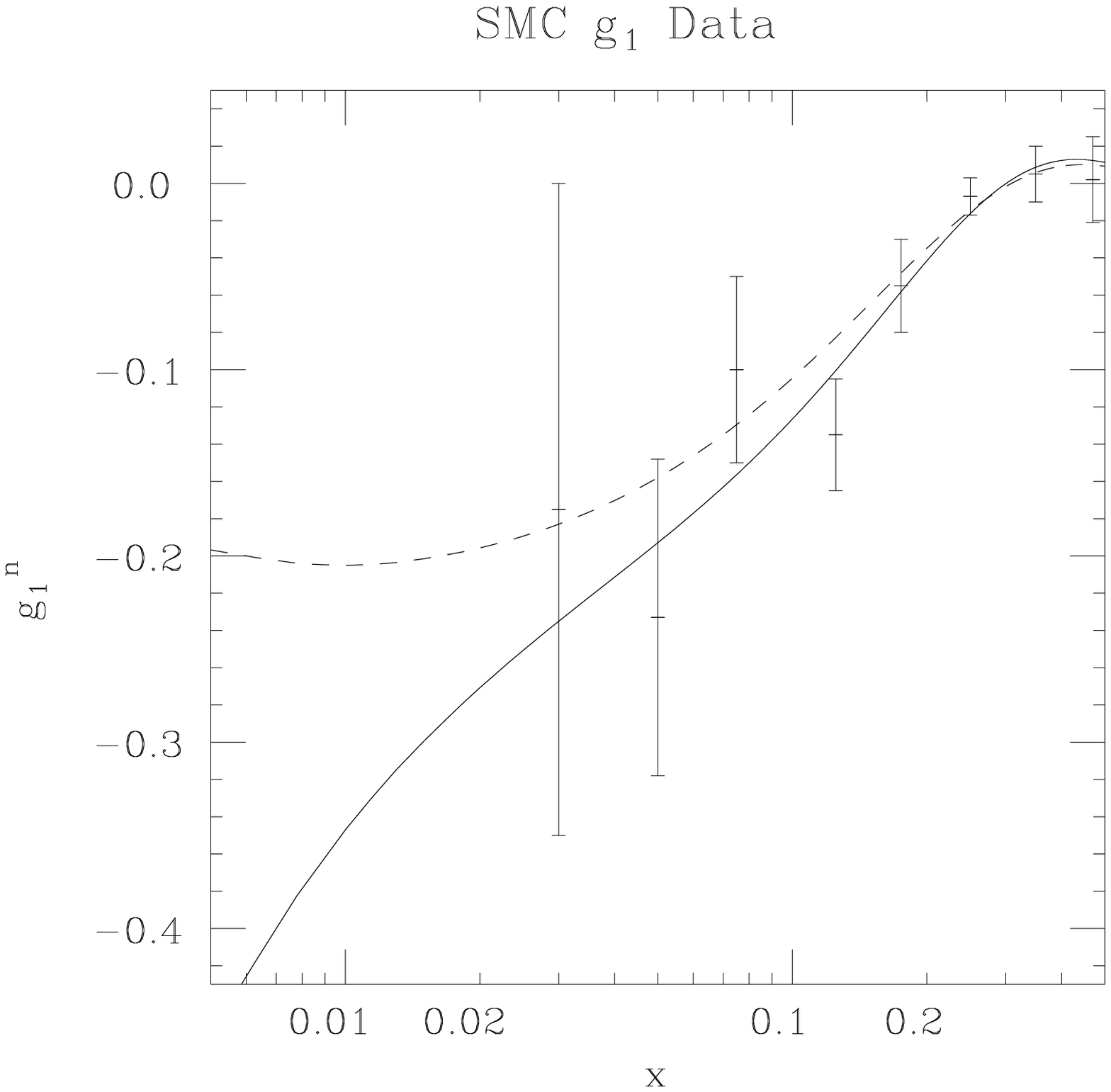}}
\caption{The $x$ dependent proton and neutron structure functions, $g_1^p$
and $g_1^n$ generated from Eqn. (71) and data. The dotted line represents the
GRV-generated distributions and the solid line, MRS-generated distributions.}
\end{figure}
\begin{figure}
{\hskip 7.0cm}\hbox{\epsfxsize7.5cm\epsffile{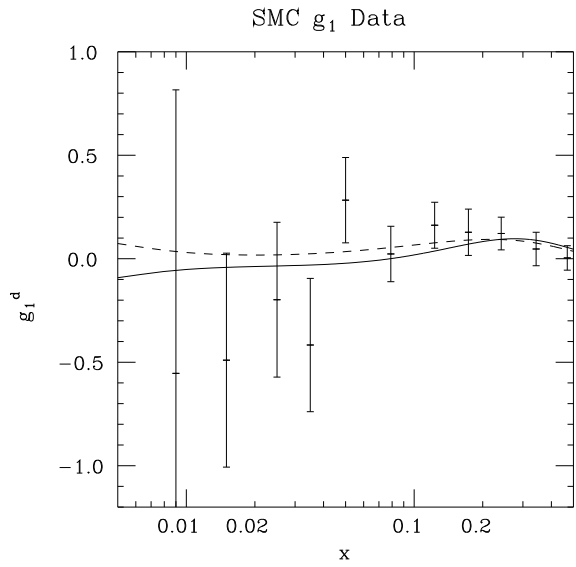}}
\caption{The $x$ dependent deuteron structure function, $g_1^d$, using the
same technique as in figure 2.}
\end{figure}

\subsubsection{Open Physics Questions}

The proceeding analyses emphasize a number of key physics questions which the
latest DIS experiments have raised. These include, but are not limited to:
\begin{itemize}
\vspace*{-2ex}
\item the size and sign of the polarized gluon distribution
\vspace*{-2ex}
\item the amout of spin carried by the strange sea
\vspace*{-2ex}
\item the light quark flavor dependence of polarization; is there an analogy
to violation of Gottfried sum rule in $\Delta u$ and $\Delta d$?
\vspace*{-2ex}
\item the connection, if any, between the unpolarized and the polarized
distributions
\vspace*{-2ex}
\item the role of the higher order corrections at low $x$ and $Q^2$-both
perturbative and higher twist (non-perturbative)
\vspace*{-2ex}
\item the role of the orbital motion, $L_z$; does it agree with the expected
asymptotic values? \cite{Lz} \\
\end{itemize}

Theoretical models which agree with present data still disagree on many of
these points. It is therefore up to the experiments to decide which of these
explain the various aspects of constituent spin contributions to the nucleons.
In the next section, we outline the various polarization experiments which
will address these issues. \\

\subsection{Future Experiments}

\subsubsection{Introduction}

There are a number of experiments which are technologically feasible that would
supply some of the missing information about these distributions. Detailed
summaries can be found in references by Ramsey \cite{RRev} and
Nurushev. \cite{Nu}
We have seen that the existing data have enabled us to formulate appropriate
questions which probe the spin properties of nucleons. However, there are
a number of questions which remain unanswered and will only be accessible with
more data at different energies and momentum transfers. Fortunately, there are
a number of experimental groups that are planning polarized beam experiments
at existing accelerators. With recent advances in polarized beam, target,
detector and accelerator technology, it is now possible to do these experiments
at higher energies and momenta in order to study the physics over a large
kinematic range.  The large average luminosities of these experiments and the
success of Siberian Snakes makes all of the following feasible. This section
will include some of the proposed experiments related to the spin structure of
nucleons, in light of the questions presented in the last section. \\

One of the ways to categorize polarization data is as follows:
\begin{itemize}
\vspace*{-2ex}
\item deep-inelastic scattering of polarized leptons (e, $\mu$) on polarized
nucleon targets ($p$, $n$, $d$)
\vspace*{-2ex}
\item  photo-production of jets in high energy polarized $ep$ colliders
\vspace*{-2ex}
\item production of pions and direct photons in polarized $pp$ and $p\bar p$
scattering
\vspace*{-2ex}
\item charmed meson production (J/$\psi$ and $\chi$) in $pp$ collisions
\vspace*{-2ex}
\item direct photon and jet production in polarized $pp$ collisions
\vspace*{-2ex}
\item  lepton pair production (Drell-Yan) in polarized processes
\vspace*{-2ex}
\item heavy baryon (hyperon) production in unpolarized $pp$ collision
\end{itemize}

All of these are designed with the measurement of particular distributions in
mind. Some can also provide crucial tests of QCD. There are a number of these
presently planned at the following locations (alphabetically): (1) CERN
(Switzerland), (2) DESY (Germany), (3) LISS (Indiana, USA), (4) RHIC
(Brookhaven, USA), (5) Serpukhov (Russia) and (6) SLAC (USA). A partial list
of experiments with their corresponding energies is given in Table V. This is
to give a general idea of the wide range of energies and kinematic regions to
be covered. \\

\begin{center}
{\bf Table V} \\

\begin{tabular}{|c||c||c||c|}
\hline
 Experiment & Location & Energy $\sqrt s (GeV/c)$ & Luminosity (cm$^{-2}$
s$^{-1}$)  \\
\hline
  HERMES  & HERA (DESY) & $30\>(e)$ on $820\>(p)$ & $2\cdot10^{31}$  \\
\hline
  SPIN   & HERA (DESY)  &  820           & $2\cdot10^{31}$  \\
\hline
  RHIC   & Brookhaven   & $60\to 500$    & $2\cdot10^{32}$  \\
\hline
  LISS   & Indiana      & 20             & $1\cdot10^{32}$  \\
\hline
  COMPASS & CERN        & 120  		 & $1\cdot10^{32}$   \\
\hline
  NEPTUN-A & Serpukhov  & 400  		 & $1\cdot10^{31}$   \\
\hline
  E155   & SLAC         &  48 (e) 	 &  ---  \\
\hline
\end{tabular}
\end{center}

In the following discussion, we will describe some of these experiments with
regard to the physics that they probe, namely:
\begin{itemize}
\vspace*{-2ex}
\item a. $\Delta q$ measurements (valence and sea)
\vspace*{-2ex}
\item b. $\Delta G$ measurement
\vspace*{-2ex}
\item c. $L_z$ determination
\vspace*{-2ex}
\item d. Nuclear measurements - $F_2^D$, $g_1^A$, $h_1^{N,D}$
\vspace*{-2ex}
\item e. higher order corrections and other tests of QCD
\end{itemize}

A partial list of the experiments for these categories is in Table VI. These
will be discussed in the following subsection.
The experiments discussed here represent a sampling of those which directly
relate to the subject of the spin structure of nucleons. Many of these are
just a small fraction of the polarization experiments which can be performed
at these accelerators, but they form an integral part of the program. \\

\begin{center}
{Table VI} \\
\begin{tabular}{|c|c|c|c|}
\hline
 Experiment & Proposed Type & Measured Quantities & Distribution \\
\hline
 HERMES     & DIS           & $A_1^p$, $g_1^p$    & $\Delta q$, $\Delta G$ \\
 E155       & DIS	    & $g_1^p$, $g_1^d$    & $\Delta q$, $\Delta g$ \\
 SPIN       & Elastic pp    & $A_N$, $A_{NN}$     & Helicity NC  \\
 RHIC/COMPASS & Charm prod. & $A^c$ 		  & $\Delta G$      \\
 RHIC (STAR) & Jet, $\pi$, $\gamma$ prod & $\Delta\sigma_L$, $A_{LL}$
& $\Delta G$ \\
 RHIC (PHENIX) & Drell-Yan  & $A^{DY}$  	  & $\Delta S$   \\
 LISS       & Inelastic    & $\sigma_L$, $\sigma_T$, $\Delta\sigma_L$
 & $\Delta G$  \\
 SLAC  	    & Charm prod.  & $A^c$		&  $\Delta G$  \\
 SLAC  	    & W$^{\pm}$ prod  & $A^{W}$		&  $\Delta q_i$  \\
 LHC  	    & W$^{\pm}$ prod  & $A^{W}$		&  $\Delta G$   \\
\hline
\end{tabular}
\end{center}

\subsubsection {Valence and Sea sensitive experiments}

Deep Inelastic Scattering: The E155 experiment has been approved at SLAC. This
experiment is designed to probe slightly smaller $x$ while greatly improving
statistics and systematical errors. With lower error bars at small $x$, the
extrapolation should achieve a more accurate value for the integrated
distributions and narrow the ranges of constituent spin contributions even
further. \\

There has been considerable discussion about performing the COMPASS polarization
experiments at the LHC at CERN. Depending on the approved experiments,
there is the possibility of probing small $x$ and doing polarized
inclusive experiments to measure both sea and gluon contributions to proton
spin. These could be made in complementary kinematic regions to those of
the other accelerators. There are tentative plans to do polarized $W^{\pm}$
production, which provides a measure of the $x$-dependent sea distributions.
Polarized $W^{\pm}$ production is also planned at SLAC and would provide useful
flavor dependent sea information in a slightly different kinematic region
than that of CERN. \\

The latest experiments at HERA in Hamburg have accelerated a large
flux of polarized electrons from the storage ring and collided them with a
gaseous target. The gaseous target has helped to eliminate some of the
systematic errors characteristic of solid targets, which were used in the
other experiments. This can be repeated for proton targets to test the BSR. 
With more events and the lower error bars at small $x$, it will be easier to
extrapolate $g_1^p(x)$ and achieve a more accurate integrated value,
$\langle g_1^p\rangle$. Thus, comparison to the Bjorken and Ellis-Jaffe
sum rules will be more accurate, as will the ability to determine the polarized
sea values from this data. \cite{BFR, Bour} \\

Anselmino, \etal, \cite{Anselmino96b} have proposed doing charged current
interactions ($l^{\pm}p\to \nu X$), which could measure various linear
combinations of flavor dependent polarized distributions. These experiments are
feasible and could put further constraints on the quark contributions to spin.
\\

Tests of the valence quark polarized distributions can be made, provided a
suitable polarized antiproton beam of sufficient intensity could be developed.
\cite{RS91} This would provide a good test of the Bjorken sum rule via
measurement of $\langle \Delta q_v\rangle$ and the assumption of a flavor
symmetric up and down sea. \\

Lepton pair production (Drell-Yan) processes provide another clean measure of
the polarized sea. \cite{Cheng90} 
The Relativistic Heavy Ion Collider (RHIC) at Brookhaven is designed to be
an accelerator of both light and heavy ions. \cite{RHIC} The high energy
community has proposed that polarized $pp$ and $p\bar p$ experiments be
performed, due to the large energy and momentum transfer ranges which should be
available. The energy range will be made in discrete steps between 50 and 500
GeV, and the momentum transfer range also covers a wide kinematic region.
There are two main proposed detectors, STAR and PHENIX, which have different
but complementing capabilities. Polarized Drell-Yan experiments are planned,
which would give reasonable estimates to the polarized sea for each flavor.
The PHENIX detector is suitable for lepton detection and the wide range of
energies and momentum transfers could yield a wealth of Drell-Yan data over a
wide kinematic range. The x-dependence of the polarized sea distributions could
then be extracted to a fair degree of accuracy. Kamal \cite{Kamal} has
calculated next-to-leading order (NLO) corrections to Drell-Yan processes and
has made predictions for RHIC energies. \\

There has been a major effort to propose a double polarized $pp$ mode at HERA,
which would accelerate polarized protons. There is a wealth of both inclusive
and exclusive experiments that can be done there at an energy range
complementary to that of the other accelerators. This would require a major
upgrade and installation of Siberian snakes, but the physics output potential
is great. This initiative would be labeled HERA-$\vec N$. For a comprehensive
review of the prospects at HERA-$\vec N$, see the proceedings
of the workshop at Zeuthen \cite{HERAN} and references by Anselmino, \etal,
\cite{Anselmino96c} and Nowak. \cite{WDN} Gehrmann and Stirling \cite{GS} have
calculated the NLO Drell-Yan asymmetries for both unpolarized and polarized
experiments at HERA-$\vec N$ energies. Combining these with the RHIC
measurements would give a good measure of the polarized sea over a large energy
range. \\
 
\subsubsection{Gluon sensitive experiments}

Knowledge of the polarized gluon distribution is important for both the anomaly
contribution to nucleon spin and to its over-all spin contribution via the $J_z$
sum rule discussed earlier. Thus, experiments sensitive to $\Delta G$ are a
high proirity item. The following is a brief discussion of the possible gluon
sensitive experiments and their corresponding proposed lacations. \\

(1) Jet production in polarized $ep$ collisions: It has been suggested that
$\Delta G$ could be measured in $ep$ collisions which produce one or two jets
of hadrons from the photon-gluon fusion process. \cite{SV, DeR} This experiment
could be done at HERA, but the experted asymmetries, even for a large polarized
gluon distribution are only at the few percent level and may be difficult to
measure with a sufficient degree of certainty to distiguish between the gluon
models. \\

(2) Jet production in $pp$ collisions: Even before the spin crisis was
popular, it was known that jet production in polarized $pp$ collisions could
be a sensitive measure of $\Delta G$. \cite{RRS}
The STAR detector at RHIC is suitable for inclusive reactions involving jet
measurements. This would provide an excellent measurement of the $Q^2$
dependence of $\Delta G$ due to the large range of energies available there.
The NLO corrections to photon plus jet production have been calculated by
Gordon. \cite{Gordon96a} The corresponding asymmetries
range from a few percent for small gluon polarization to about 30\% at
HERA energies, if $\Delta G$ is large. Should DESY proceed with plans to
polarize their proton beam, this experiment could be performed there,
complementing the kinematic regions covered by RHIC and CERN. \\

(3) Direct photon and double photon production: A clean signature for $\Delta
G$, but one that is harder to measure, is that of direct photon production,
($\vec {p}\vec{p}\to \gamma\>+\>X$). \cite{QRRS, BeQ}
The NLO corrections have been calculated for HERA-$\vec N$ \cite{CM, GV} and 
RHIC. \cite{Gordon96b} Although the asymmetries are only very large for the
largest gluon models, the signal is clean and could distinguish if $\Delta G$
is large. Numerical simulations have been done in NLO for the double photon
production processes ($\vec {p}\vec{p}\to \gamma\gamma\>+\>X$). \cite{CG}
 At RHIC energies, the asymmetries can be from a few percent to
20\% for the larger $\Delta G$, but the cross sections arequite small for
the optimal momenta (picobarns or smaller). Thus, it is unclear whether this
is viable unless very high luminosities can be reached. \\

(4) Charm production in polarized collisions are also sensitive to $\Delta
G$ and should be performed at both RHIC and HERA-$\vec N$. The double spin
assymetries for J/$\psi$ production have been calculated and are sensitive
to the color-octet contribution. \cite{G, TerT} These asymmetries are only on
the few to 10\% level. The two-spin asymmetries for $\chi$ production have also
been calculated. \cite{Morii} These asymmetries are quite small unless
the polarized gluon distribution is quite large ($\langle \Delta G\rangle
\approx 6$). Thus, $\chi$ production could distinguish an extremely large
distribution from other models. Open charm production from lepton-hadron
scattering has been proposed at LHC by the COMPASS group. \cite{HERMES, BoL, NH}
This provides another method to measure
the photon-gluon fusion process and has very small uncertainties associated
with the proposed experiment. These asymmetries can be quite large and provide
a good potential for extending our knowledge of $\Delta G$. \\
           
(5) Inclusive reactions involving pion production would be alternate tests of
the $Q^2$ dependence of $\Delta G$. \cite{RS91} These could be
done at any of the aforementioned accelerators and would provide a good cross
check of measurements of $\Delta G$ in similar kinematic regions. \\

\subsubsection {Experiments Probing Higher order QCD Effects}

There are other polarized experiments which would provide tests of QCD and
give a measure of some of the higher twist effects which were previously
discussed. \\

(1) Elastic scattering, especially at a large $-t$ range, could shed light on
both helicity non-conservation at the hadronic level and non-perturbative
long range effects. Elastic experiments have been proposed for HERA, RAMPEX
at Serpukhov and LISS in Indiana. \cite{RS93} \\

(2) Recently, a proposal for a new light ion accelerator was announced, which
will specialize in polarization experiments. \cite{CV} The
Light Ion Spin Synchrotron (LISS) would be located in Indiana to perform a
variety of polarization experiments for both high energy and nuclear physics.
The energy range would be lower that most other experiments, thus complementing
the kinematic areas covered. Furthermore, both proton and deuteron beams
could be available to perform inclusive scattering experiments. They propose
to measure longitudinal and transverse cross sections and spin asymmetries,
which will address the normalization of the proton wave function. Elastic
scattering measurements of $A_N$ at moderate momentum transfer $-t$ could give
valuable information regarding helicity non-conservation in this region. \\

(3) Measurement of the transverse spin and transversity distributions would 
probe higher-twist non-perturbative effects as well. \cite{JS}
The higher twist parton distributions could also be measured at RHIC and
HERA. \cite{Kir, Lad, Kha} \\

(4) Single spin asymmetries in DIS (where only one of the scatterers is
polarized) have been calculated by many groups. These provide helicity
conservation tests as well as measures of higher twist processes.
\cite{Anselmino96d, TT, Berlin} These experiments could be done at HERA. The
$pp$ single spin asymmetries can also be done at HERA, RHIC and the LHC. These
are good measurments of higher twist contributions to the structure functions.
The $pp$ single spin experiments are also planned by the RAMPEX collaboration
(Russian-AMerican Polarization EXperiment). \cite{RAMPEX} \\

This is a summary of some of the key experiments which will answer the
questions posed throughout this review. There are many other important
polarization experiments which test QCD and non-standard physics, but they will
not be covered here. However, from this list alone, it is easily seen that
polarization experiments are crucial to our understanding of the most
fundamental properties of the elementary particles. \\

\subsection{Conclusion}

This review has outlined the key theoretical and experimental elements which
have contributed to our understanding of the constituents' contribution to the
spin of nucleons. Analyses of existing data indicate that the spin structure
of nucleons is non-trivial and has led to the formulation of a crucial set of
questions to be answered about this structure. The key remaining questions are
related to the strange sea and gluon polarizations. The experiments discussed
here can be performed in order to shed light on these questions. In performing
these experiments, an added benefit is that crucial tests of QCD and the quark
model will be simultaneously be made. The questions of factorization and
validity of the sum rules are among the questions which can be addressed.
In the past few years, theorists have made considerable progress in calculating
higher order corrections to the appropriate sum rules and structure functions
and experimentalists have added more accurate and comprehensive DIS data.
This has contributed to narrowing the range of possible spin contributions from
the constituents, but has also raised other questions about spin structure of
nucleons. \\

There is still much work to be done in constructing a suitable model for
spin transfer among constituents, along with calculating higher order
corrections to the various spin processes which can be used to test the models.
There are many experiments planned at various locations, which will create
work for the experimentalists and phenomenologists. These suggestions do not
include some of the other areas of probing spin phenomena, mentioned at the
beginning of this review. These other areas cover the spectrum of both high
energy and medium energy nuclear physics as well. Thus, we are in an
interesting period of spin physics. One where considerable progress has been
made, only to discover that there is much more to be done to increase our
understanding of the fundamental nature of matter through the use of
polarization. \\

{\bf References} \\

Other Review papers with related but complementary information: \\

\noindent
M. Anselmino, \etal, Phys. Rep. \rf{261}{1}{95} \\
D. Bass and A. W. Thomas, Prog. Part. Nucl. Phys., {\bf 33}, (1994) \\
H.-Y. Cheng, Int. J. Mod. Phys., \rf{A11}{5109}{96} (also hep-ph/9607254) \\
J. Ellis and M. Karliner, hep-ph/9601280 (CERN-TH/95-334; TAUP-2316-96) \\
B. Frois, Prog. Part. Nucl. Phys., {\bf 34}, (1995) \\
I. Hinchcliffe and A. Kwiatkowski, hep-ph/9604210 (LBL-38549) \\
S.E. Troshin and N.E. Tyurin, Particle World \rf{3}{165}{93}. \\

References cited in paper:

\end{document}